\documentclass[twocolumn]{aa}
\usepackage{graphicx}
\usepackage{natbib}
\usepackage{txfonts}
\begin{document}
   \title{Size and Properties of the NLR in the Seyfert--2 Galaxy
   \object{NGC\,1386}\thanks{Based on observations made with ESO Telescopes at
   the Cerro Paranal Observatory under programme ID 72.B--0144}}

   \author{Nicola Bennert\inst{1}
          \and
           Bruno Jungwiert\inst{2,3} \and Stefanie Komossa\inst{4} \and
          Martin Haas\inst{1} \and Rolf Chini\inst{1}
          }

   \offprints{Nicola Bennert}

   \institute{Astronomisches Institut Ruhr--Universit\"at Bochum,
              Universit\"atsstrasse 150, D--44780 Bochum, Germany
              \email{nbennert@astro.rub.de}, \email{haas@astro.rub.de}, \email{chini@astro.rub.de}
\and
Astronomical Institute, Academy of Sciences of the Czech Republic,
Bo{\v c}n\'\i\ II 1401, 141 31 Prague 4, Czech Republic, 
\email{bruno@ig.cas.cz}
          \and
             CRAL--Observatoire de Lyon, 9 avenue Charles Andr{\'e}, F--69561 
Saint--Genis--Laval cedex, France
\and
   Max--Planck Institut f\"ur extraterrestrische Physik,
              Giessenbachstrasse 1, D--85748 Garching, Germany \email{skomossa@xray.mpe.mpg.de}
}

   \date{Received; accepted}

   \abstract{We study the narrow--line region (NLR) 
of the Seyfert--2 galaxy \object{NGC\,1386}
by means of long--slit spectroscopy obtained with FORS1 at the VLT.
We use the galaxy itself for subtracting the stellar template, applying
reddening corrections to fit the stellar template to the spectra of the
NLR. The continuum gets steadily redder towards the nucleus. The spatial
distribution of the reddening derived from the Balmer decrement differs from the
continuum reddening, indicating dust within the NLR with a 
varying column density along the
line of sight. Using spatially resolved spectral diagnostics,
we find a transition between central line ratios falling into the
AGN regime and outer ones 
in the \ion{H}{ii}--region regime, occuring at a radius of $\sim$6\arcsec~(310
pc) in all three diagnostic diagrams. 
Applying \texttt{CLOUDY} photoionisation models, we show that the observed distinction
between \ion{H}{ii}--like and AGN--like ratios in \object{NGC\,1386} represents a true
difference in ionisation source and cannot
be explained by variations of physical parameters such as
ionisation parameter, electron density or metallicity. We 
interpret it as a real border between the NLR,
i.e. the central AGN--photoionised region and 
surrounding \ion{H}{ii} regions.
We derive surface brightness, 
electron density, and ionisation parameter
as a function of distance from the
nucleus. Both the electron density and the ionisation parameter decrease with
radius.  We discuss the consequences of these observations
for the interpretation of the empirical NLR size--luminosity relation.
In the outer part of the NLR, we find evidence for shocks, resulting in a
secondary peak of the electron--density and ionisation--parameter distribution 
north of the nucleus. We compare the NLR velocity curve with the stellar one
and discuss the differences.
}
\titlerunning{The NLR in \object{NGC\,1386}}
\authorrunning{N. Bennert et\,al.}

   \maketitle
%
%________________________________________________________________

\section{Introduction}
\subsection{The narrow--line region in active galaxies}
The narrow--line region (NLR) in active galactic nuclei (AGNs)
is thought to be ionised by the luminous central engine, most likely an accreting
supermassive black hole (BH). The narrow emission--lines are used in
essentially all optical surveys to find and classify AGNs. Emission--line
diagnostic provides a powerful tool to study the size and properties of the NLR.
However, until today, even basic issues remain open: For example, what
determines the NLR size and structure? 
What are the physical conditions within the NLR such as electron
density and ionisation parameter and do they vary with distance from the nucleus?

Recently, a relation between NLR size and [\ion{O}{iii}]\,$\lambda$5007\,\AA~(hereafter 
[\ion{O}{iii}]) luminosity has been discovered from HST narrow--band images of 
seven radio--quiet PG quasars \citep{ben02}. However, the slope is discussed
controversially: While \citet{ben02} find $R \propto L_{\rm [OIII]}^{0.5}$,
suggesting a self--regulating mechanism that determines the size to scale with
the ionisation parameter, 
\citet{sch03b} report a relation of $R \propto L_{\rm [OIII]}^{0.33}$ for their
sample of 60 Seyfert galaxies imaged with HST. Such a slope is
expected for gas ionised by a central source (Str\"omgren) in the case of 
constant density.

Studying the [\ion{O}{iii}] emission alone as a tracer
of the NLR size and structure has the major drawback that this
emission can be contaminated by contributions from
starbursts, shock-ionised gas or tidal tails, resulting in an apparent
increase of the NLR. In addition, different sensitivities can lead to 
different size measurements: When comparing
groundbased [\ion{O}{iii}] images of Seyfert galaxies
from \citet{mul96a} with the HST snapshot survey of \citet{sch03a},
the latter reveal, on average, six times smaller NLR sizes, probably due to 
the 15 to 20 times lower sensitivity.
These considerations question the definition ``NLR size'' from 
[\ion{O}{iii}] imaging alone.

Long--slit spectroscopy
is a valuable alternative approach as it can 
directly probe the size in terms of AGN photoionisation and discriminate
the contribution of stellar or shock ionisation. In addition,
several physical properties of the NLR such as the electron density
and the ionisation parameter can be measured 
as a function of distance from the nucleus. Thus, it allows e.g. to probe the constant
density needed to explain the 0.33 slope of the NLR size--luminosity relation.
These parameters are also of
general interest as they are important input parameters of photoionisation
models of the NLR.

A well suited object for such an approach is the nearby 
($v_{\rm hel}$ = 868 $\pm$ 5 km\,s$^{-1}$, 
NED\footnote{NASA/IPAC Extragalactic Database}) Seyfert--2
galaxy \object{NGC\,1386}, allowing for a detailed study of the extended
NLR down to small spatial scales (1\arcsec~$\simeq$ 52 pc).

We here describe long--slit spectroscopic 
observations obtained with FORS1\footnote{FOcal Reducer/low
dispersion Spectrograph} at the VLT\footnote{Very Large Telescope, Cerro Paranal,
Chile (ESO)}
and the methods used to determine the ``real'' NLR size 
as well as physical conditions within the NLR. These
methods were applied to a larger 
sample of Seyfert galaxies, revealing similar results as those for \object{NGC\,1386}
which will be discussed in a subsequent paper.

\subsection{\object{NGC\,1386}}
\label{literature}
\object{NGC\,1386} is one of the nearest known Seyfert galaxies
($D \sim 11$ Mpc for $v_{\rm 3K}$ = 774 km\,s$^{-1}$ 
and H$_0$ = 71 km\,s$^{-1}$\, Mpc$^{-1}$;
1\arcsec~$\sim$ 52 pc). 

Due to its relatively high inclination ($i \simeq 71\degr$),
the morphological galaxy type has been a matter of debate: While
several authors favour S0, an Sa classification is assumed by \citet{wea91}. 
\citet{mal98} give a morphological type of Sb/c
based on HST imaging, noting that they possibly have missed bars on larger
spatial scales. In RC3 \citep{vau91}, \object{NGC\,1386} is indeed classified as barred
SB0 galaxy. Also NED gives SB(s)0+ as morphological classification.

\object{NGC\,1386} has been investigated by various authors.
We here summarise the most important results with respect
to our study.

\citet{wea91} study the extended NLR by means
of groundbased imaging as well as long--slit spectroscopy with multiple slit positions.
They show a 2D coverage of the inner kpc of this galaxy
and discuss the overall morphology, velocity field, 
electron--density distribution, and ionisation structure.
They use the galaxy itself as stellar template to correct for underlying
absorption lines. Using one diagnostic line--ratio diagram
([\ion{O}{iii}]/H$\beta$ versus [\ion{N}{ii}]\,$\lambda$6583\,\AA/H$\alpha$),
they determine the NLR (i.e. the AGN powered region)
to extend $\sim$6\arcsec~to the north and south
of the nucleus while emission lines further out can be attributed
to \ion{H}{ii} regions.
They find an electron density which is decreasing with radius.
The velocity field is interpreted in terms of a combination of a normally
rotating component and a component undergoing high--velocity infall or outflow.

\citet{mau92} resolve the NLR down to a linear scale of 0\farcs3 ($\simeq$ 15 pc) using 
speckle interferometry and detect individual NLR clouds on these scales.
They argue that the ionising radiation must be absorbed on scales $<$15 pc
in a clumpy structure of the NLR.

\citet{tsv95} give positions and fluxes of 44 \ion{H}{ii} regions 
distributed in a ring seen in their 
continuum--subtracted
H$\alpha$+[\ion{N}{ii}]\,$\lambda$$\lambda$6548,6583\,\AA~(hereafter
[\ion{N}{ii}]) image.
The distances of the \ion{H}{ii} regions are $\sim$6--12\arcsec~from
the nucleus at a position angle (p.a.) of the major axis of
the ring of 25\degr,
comparable to the p.a. of the major axis of the galaxy [25\degr; RC3
\citep{vau91}]. 

\citet{fer00} present HST images of the [\ion{O}{iii}] and H$\alpha$+[\ion{N}{ii}] emission.
The images show the presence of very strong dust features, especially on the
north--west side of the galaxy and in the nuclear regions. They find gradients
in the [\ion{O}{iii}]/([\ion{N}{ii}] + H$\alpha$) ratio which at least partly
cannot be due to dust lanes but may corres\-pond to a real transverse change in
the excitation conditions of the ionised gas.
\citet{fer00} report a faint inclined ring of \ion{H}{ii}--region emission
extending out to
$\sim$12\arcsec~from the nucleus with a p.a. of 25\degr,
in agreement with the results from \citet{tsv95}.
The [\ion{O}{iii}]
image of \citet{fer00} is included in the study of \citet{sch03a} and presented here
in Fig.~\ref{slit} with the slit position of our observation
overlayed (p.a.$_{\rm [OIII]}$ = 5\degr). It traces the direction of the maximum extent in
[\ion{O}{iii}] which is displaced by 20\degr~from the host galaxy major axis
(p.a.$_{\rm galaxy}$ = 25\degr) 
and by $\sim$15\degr~from the extended radio emission [p.a.$_{\rm radio}$ = 
170\degr, \citet{nag99}].

\cite{ros00} and \citet{sch03} study the kinematics of the inner emission--line
region in detail. Both find evidence for a regular velocity field
showing signs of rotation. The line profiles show pronounced substructure
suggestive of non--circular motions like locally expanding gas systems.
\citet{sch03} suggest a near--edge--on warped rotating spiral disk
as traced by H$\alpha$ with a central velocity gradient corresponding
to a Keplerian mass estimate of $\sim 5 \cdot 10^9$ M$_{\odot}$ inside
0.8 kpc. Like \citet{mau92}, \cite{ros00} detect several
individual NLR components.

\object{NGC\,1386} is also included in the sample of 18 Seyfert--2 galaxies studied
with long--slit spectroscopy by \citet{fra03}. They present reddening
and surface--brightness distributions as well as a decreasing
electron density observed along a p.a. of 169\degr. However, their data have a
significant lower signal--to--noise ratio (S/N)
than our data and, moreover, they
did not take into account the
underlying absorption owing to the contribution of the stellar population
which we show to be important.

\section{Observations}
\begin{figure}[h!]
\centering
\includegraphics[width=8.3cm,angle=-90]{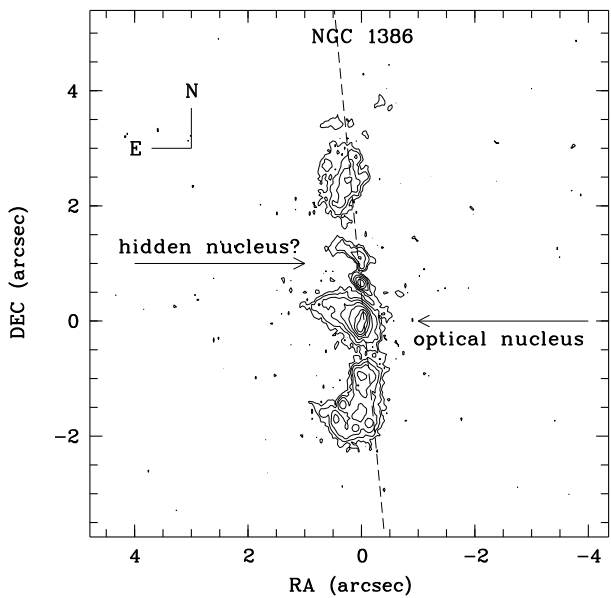}
\caption{\label{slit} [\ion{O}{iii}] continuum--subtracted image
of \object{NGC\,1386} taken with HST, kindly provided by Henrique Schmitt. 
The contours start at the 3$\sigma$ level above the background 
(2.1 $\cdot$ 10$^{-14}$\,erg\,cm$^{-2}$\,s$^{-1}$\,arcsec$^{-2}$) and increase in
powers of 2 times 3$\sigma$ (3$\sigma$ $\times$ $2^n$). 
The exposure time is 800\,s. The pixel size is
0\farcs0455. The [\ion{O}{iii}] emission was centered
on the optical nucleus determined from the
continuum image by \citet{sch03a}. The velocity curve suggests a hidden nucleus
at $\sim$1\arcsec~north of the optical nucleus. The slit position of our
observations at p.a. = 5\degr, tracing the major axis of the [\ion{O}{iii}] extension, is
indicated by the dashed line.
}
\end{figure}

The high S/N long--slit spectra of \object{NGC\,1386}
were obtained using  FORS1 attached to the Cassegrain focus of UT1 at
the VLT on the 25th of February 2004.
Observations were made in the spectral range 3050--8300\,\AA~through 
the nucleus of \object{NGC\,1386}  with exposure times of 1800\,s (to study the emission
lines far from the nucleus) and 30\,s (to gain the central emission lines which
were saturated in the 1800\,s exposure) with a
typical seeing of $\sim$1\arcsec, under the use of the atmospheric dispersion
corrector. 
The detector used is a 2048 $\times$ 2048 pixels Tektronix CCD
with 24\,$\mu$m wide square pixels.
The spatial resolution element is 0\farcs2\,pix$^{-1}$.
The slit width corresponds to 0\farcs7 on the sky projecting
to a spectral resolution of $\sim$8\,\AA~($\sim$450 km\,s$^{-1}$) 
as is confirmed by the full--width--at--half--maximum (FWHM)
of wavelength calibration lines and the [\ion{O}{i}]\,$\lambda$5577\,\AA~night--sky line.
The long slit was orientated along the p.a. 
of the maximum extent of the high excitation
gas observed in the [\ion{O}{iii}] image (p.a.$_{\rm [OIII]}$ = +5\degr; Fig.~\ref{slit})
and corresponds to 6\farcm8 in the sky.
\section{Reduction and Analysis}
\subsection{Data reduction}
\label{data}
Standard reduction including bias subtraction, flat--field correction,
and cosmic--ray rejection
was performed using the ESO
\texttt{MIDAS}\footnote{Munich Image Data Analysis System,
trade mark of ESO} software (version Nov. 99).
Night--sky spectra at 1\arcmin--3\arcmin~distance 
on both sides of any notable galaxy emission
were interpolated in the region of the galactic spectrum and subtracted.  
The spectra were rebinned 
to a scale of 2.65\,\AA~pix$^{-1}$ during wavelength calibration.
The curve of \citet{tug97} was used to correct for atmospheric extinction.
The spectra were flux calibrated using the standard star CD--32\degr9927.
Foreground Milky Way reddening was corrected using values from
\citet{sch98} as listed in NED ($E_{(B-V),G}$ = 0.012\,mag)
and the extinction law from \citet{sav79}.
Forbidden--line wavelengths were taken from \citet{bow60}. A heliocentric
correction of $-$28.8\,km\,s$^{-1}$ was added to the observed velocities. 

In the spatial direction perpendicular to the dispersion axis,
five pixel rows were averaged from the frames cleaned in this way 
according to the seeing to enhance the S/N without loosing any spatial information.
Thus, each resulting extracted spectrum of the spatially resolved emission
corresponds to 1\arcsec~and 0\farcs7 along and perpendicular
to the slit direction, respectively. 
We choose the spectrum with the maximum intensity of the continuum
as ``photometrical center'' (``zero'' on the spatial scale).
It coincides with the highest emission--line fluxes in H$\alpha$ and [\ion{O}{iii}].
In the following, we also refer to it as 
``central spectrum''.
\subsection{Subtracting the stellar population}
\label{stellarpop}
Starlight constitutes a substantial fraction of light gathered in the
optical spectra of AGNs, particularly in low--luminosity objects such
as LINERs and Seyfert galaxies. 
The emission--line fluxes are affected by underlying absorption lines
which can be large especially in the
case of H$\beta$. As a consequence, starlight
can have a great influence on the determination of the reddening
and the AGN diagnostic line ratio
H$\beta$/[\ion{O}{iii}]. Thus, removing the contribution of the
stellar population is one of the first and most critical steps in the
analysis of AGN emission--line spectra. 
The most widely used technique for starlight
subtraction makes use of an appropriately chosen spectrum of a
normal galaxy. 
However, the stellar populations in the inner regions of AGNs
present a variety of characteristics \citep{cid98}, 
giving rise to the need of a library with a large diversity of stellar
populations. Alternatively,
the shape and strength of the starlight component can be estimated by
means of stellar population synthesis techniques,
following the work pioneered by \citet{bic88}. The
classification refers to the library of synthetic
spectra of \citet{bic88}, i.e.~templates composed of different percentages
of star cluster spectra of several ages and metallicities.
An S3 template, for example, contains only the
contribution of stars older than 10 Gyr. In S4, 95\% of the light at
5870\,\AA~comes from old stars and 5\% from young and intermediate age ones.
In S5, the contribution is 85\% from old stars and 15\% from young and
intermediate age ones, etc. Thus, a low number ``Sx'' indicates a rather old 
stellar population, a high number a young one.
This method, however, involves
several assumptions of stellar modeling and does not necessarily
yield an unambiguous solution.  

Here, we applied the straightforward approach
of using the galaxy {\it itself} as starlight template which should give
the best representation of the stellar population.
The high--sensitivity spectrum allows us to average
several pixel rows with high S/N in the outer part of the galaxy spectrum. 
One assumption is that
the stellar population does not vary much in the inner few arseconds,
i.e. the bulge, as is confirmed by the close match of the stellar template 
and the absorption lines in all studied spectra.
Furthermore, we assume that
the contribution of the AGN featureless continuum is negligible
to not overestimate the underlying absorption lines by scaling
the template to the continuum in each spectrum. 
This seems to be justified for type--2 AGNs:
 \citet{cid98} find little or no dilution of the stellar lines by an
underlying featureless continuum for most of the 20
Seyfert--2s in their sample, including \object{NGC\,1386}.
\citet{wea91} apply a similar method of subtracting the stellar contribution
in \object{NGC\,1386}.

\begin{figure}[h!]
\centering
\resizebox{\hsize}{!}{\includegraphics[angle=-90]{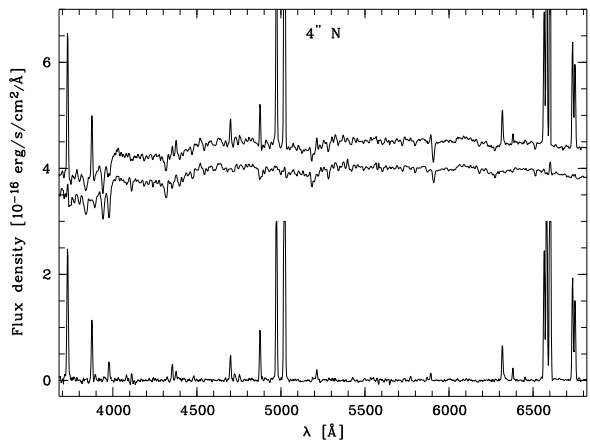}}
\caption{\label{template}
Spectrum
at a distance of 4\arcsec~north of
the nucleus. The upper
spectrum shows strong \ion{Ca}{ii} H+K absorption lines and
Balmer features. A template spectrum
of the stellar contribution derived from the galaxy  itself at
18\arcsec~north of the nucleus
was subtracted (middle spectrum). The difference spectrum is the lower one
(with the stronger emission lines truncated). Both upper spectra
are shifted by an arbitrary amount for comparison.}
\end{figure}

The stellar template was 
gained at a distance of $\sim$18$\arcsec$~(940\,pc) north from the nucleus,
averaged over 5\arcsec~and median filtered along the spectral axis over three
pixels to increase the S/N.
The extracted region is both far enough
from the central region to exclude any NLR emission,
as well as near enough to still guarantee a comparable
stellar population with high S/N.
This assumption is strenghtened by the stellar population analysis
carried out for \object{NGC\,1386} by \citet{cid98}:
The equivalent widths at the nucleus are typical of an S3 template out to
10\arcsec~where the stellar population resembles S4--S5 due to the presence
of spiral arms. This coincides with the inclined ring of emission 
seen in the H$\alpha$+[\ion{N}{ii}] image by \citet{fer00}.
Further out, the stellar population is again S3,
and we can thus use the template determined at $\sim$ 18\arcsec~to fit the
nuclear spectra out to 10\arcsec.
The resulting template spectrum is clearly dominated by absorption lines  
showing strong \ion{Ca}{ii} H+K lines as well as the Balmer series
(Fig.~\ref{template}, middle spectrum).

The template was scaled to each row of the NLR spectrum by normalization in
the red ($\sim$5400--5700\,\AA), 
justified by the fact that the slope at $\lambda \ge 5400$\,\AA~does 
not change significantly for different stellar populations \citep{bic86}.
(Note that we also chose this range as it does not cover any strong NLR emission lines.)
To allow for a possible reddening difference 
of the template and each  NLR spectrum
due to different dust amounts in different galactic regions,
we applied a reddening correction to the template
by fitting the continuum slope of the template to each NLR
spectrum [\texttt{MIDAS} command ``extinct/long''
with extinction--law from \citet{sav79}].
This procedure was often necessary to avoid a template continuum exceeding
that of the NLR continuum at some wavelength after normalization.

\begin{figure}[h!]
\centering
 \resizebox{\hsize}{!}{\includegraphics[angle=-90]{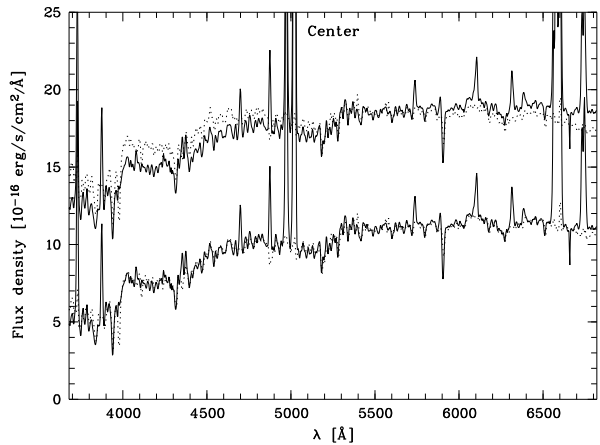}}
\caption{\label{slope} 
Central spectrum of \object{NGC\,1386} with stellar template
overlayed as dotted line. In the upper spectra, which were
shifted by an arbitrary amount for comparison, the reddening of the nucleus
with respect to the template can be clearly seen. In the lower spectra, a
reddening correction was applied to the
template to improve the match of the template and the continuum of the
nuclear spectrum.}
\end{figure}

In Fig.~\ref{slope}, we show the template with
and without reddening correction with respect to the nucleus.
The scaled (de--) reddened template was 
subtracted to gain the pure emission--line spectra.
Finally, any remaining AGN continuum contribution was fit interactively by a
continuum slope and subtracted (Fig.~\ref{template}, lower spectrum).
\subsection{Emission--line fluxes and reddening}
The fluxes of the pure emission--line spectra
were measured as a function
of distance from the nucleus by integrating along a Gaussian
fit to the line profile.
The fit routine ``\texttt{fit/spec}'' \citep{rou92} was used for this purpose.
In the three central spectra (center $\pm$1\arcsec), 
the strongest emission lines were saturated
in the 1800\,s exposure and all fluxes were derived from the 30\,s nuclear exposure.

The uncertainties in deriving the fluxes were mostly caused by uncertainties in the
placement of the continuum and were thus estimated as the product of the
FWHM of the line and the root--mean square
deviation of the local continuum fluxes. Gaussian error propagation was used
to calculate the errors of subsequent parameters such as line ratios,
ionisation parameter, etc. The resulting erros are in the range of $\sim$ 1--15\%.
Note that we did not take into account uncertainties from stellar absorption
correction and the quality of the Gaussian fits which was very good given
  the low spectral resolution of our data.

The spectra were dereddened using the recombination
value for the intensity ratio H$\alpha$/H$\beta$ = 2.87
(a typical value for $T$ = 10000 K, \citet{ost89}, Table 4.2) and an average
reddening curve (\citet{ost89}, Table 7.2).
Note that in the following, only those spectra are used 
which have emission--line fluxes exceeding the S/N ratio of 3.

\section{Results and Discussion}
\begin{figure}[h!]
\centering
 \resizebox{\hsize}{!}{\includegraphics[angle=-90]{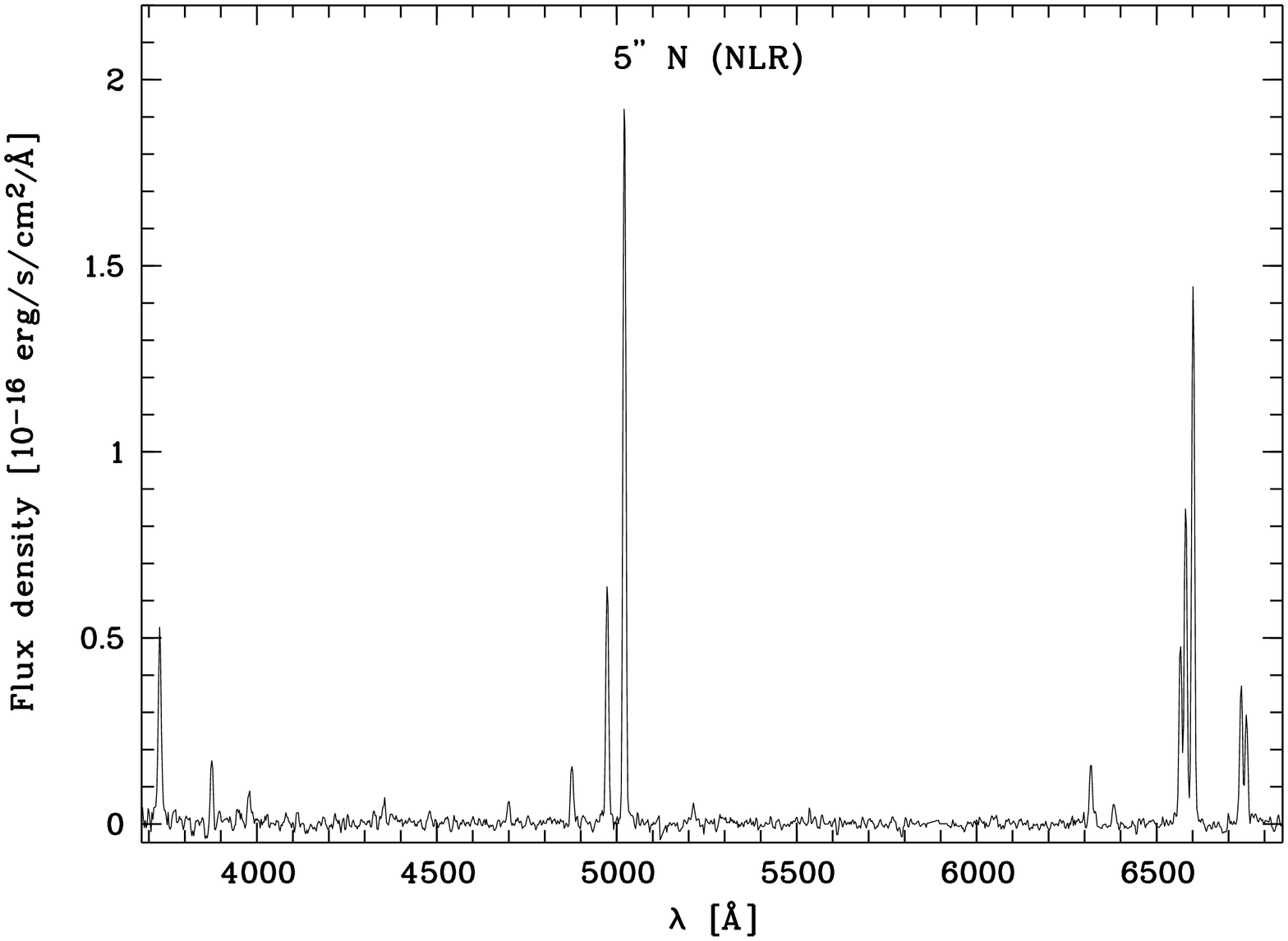}}
 \resizebox{\hsize}{!}{\includegraphics[angle=-90]{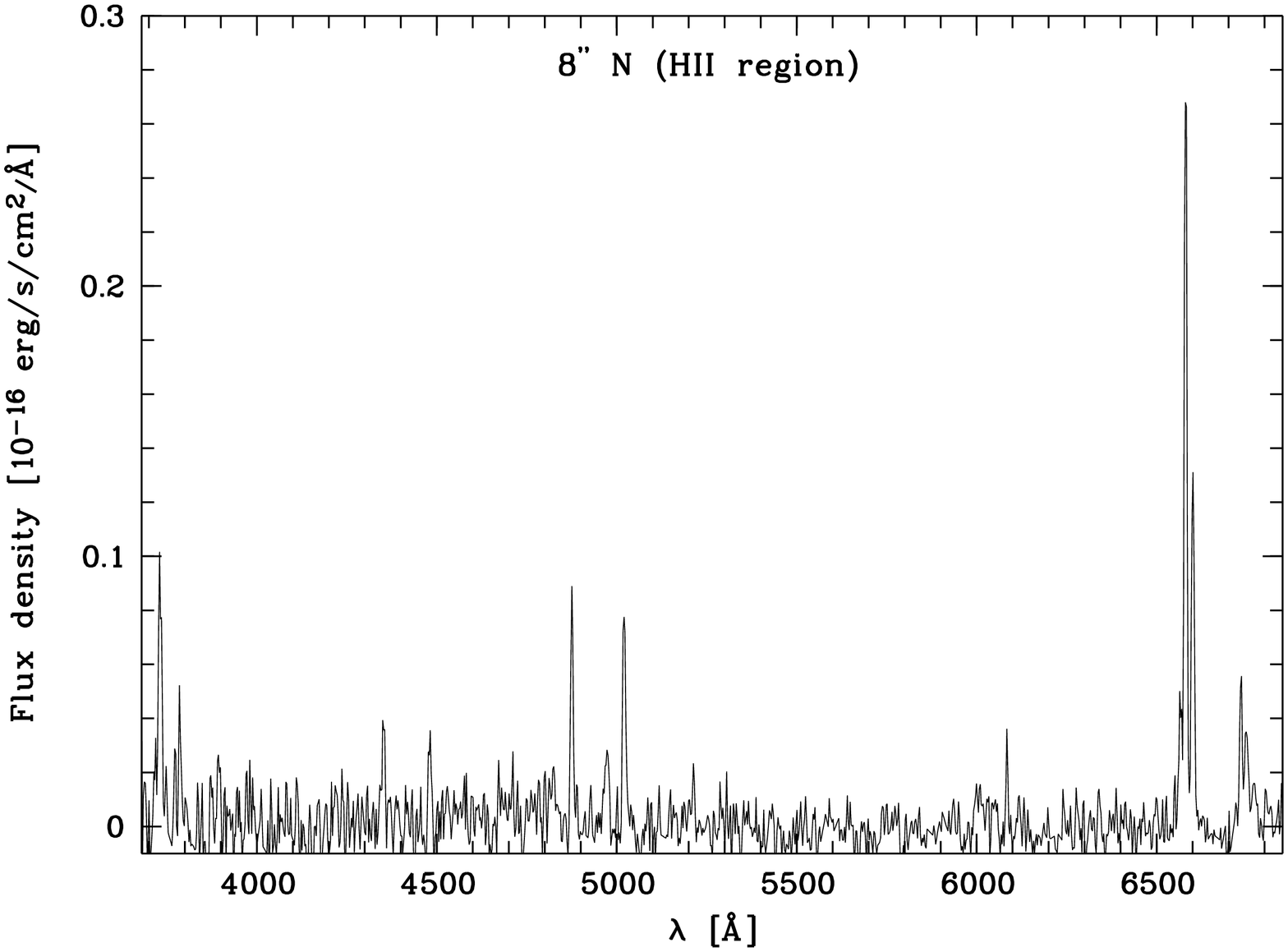}}
\caption{\label{spectra} 
Spectra of \object{NGC\,1386} at a distance of 5\arcsec~north of
the nucleus ({\it upper panel}) as well as 8\arcsec~north of
the nucleus ({\it lower panel}), respectively.
While at 5\arcsec~north of
the nucleus, the lines show typical AGN ratios (i.e. [\ion{O}{iii}]/H$\beta$
$>$ 3, [\ion{N}{ii}]/H$\alpha$ $>$ 1), at 8\arcsec~distance, the ratios are
typical for \ion{H}{ii} regions 
(i.e. [\ion{O}{iii}]/H$\beta$ $<$ 3, [\ion{N}{ii}]/H$\alpha$ $<$ 1).
}
\end{figure}

While a spectrum of \object{NGC\,1386} at 4\arcsec~north of the nucleus as well as a
nuclear spectrum were already presented in Figs.~\ref{template} and~\ref{slope},
Fig.~\ref{spectra} additionally shows representative spectra at 
5\arcsec~and 8\arcsec~north of the nucleus 
to demonstrate the emission--line variations  with distance from the nucleus.

In Table~\ref{lineratio}, we present both the observed and
reddening--corrected line--intensity ratios relative to H$\beta$ from the
nuclear spectrum (uncorrected for slit losses). 
For pairs of lines ([\ion{O}{iii}], [\ion{O}{i}], and [\ion{N}{ii}]) with a fixed
line ratio ($\sim$3:1), only the brighter line is listed.
The reddening--corrected H$\beta$ luminosity is (1.42$\pm$0.01) $\cdot 10^{40}$ erg\,s$^{-1}$.

From the [\ion{O}{iii}]($\lambda$4959\,\AA+$\lambda$5007\,\AA)/$\lambda$4363\,\AA~emission--line 
ratio, we determine the electron temperature $T_e$ in the nuclear
spectrum to $T_e = 15650 \pm 1500$\,K.

\begin{table}
\begin{minipage}[t]{\columnwidth}
\begin{center}
\caption[]{\label{lineratio}
Observed and reddening--corrected 
line--intensity ratios relative
to H$\beta$ for the nuclear spectrum.}
\begin{tabular}{lcc}
\hline
\hline
Line & \multicolumn{1}{c}{$F_{\rm obs}$} & \multicolumn{1}{c}{$F_{\rm dered}$} \\
\hline
$[\ion{O}{ii}]\,\lambda3727$\,\AA & 1.81  $\pm$ 0.02 & 2.63\\
$[\ion{Ne}{iii}]\,\lambda3869$\,\AA & 0.77  $\pm$ 0.04 & 1.07\\
H$\epsilon+[\ion{Ne}{iii}]\,\lambda$3967\,\AA & 0.48  $\pm$ 0.05 & 0.64\\
$[\ion{O}{iii}]\,\lambda$4363\,\AA & 0.19  $\pm$ 0.02 & 0.23\\
$\ion{He}{ii}\,\lambda$4686\,\AA & 0.46  $\pm$ 0.05 & 0.48\\
$[\ion{O}{iii}]\,\lambda$5007\,\AA & \hspace*{-0.3cm}11.34  $\pm$ 0.1 & \hspace*{-0.15cm}10.73\\
$[\ion{Fe}{vii}]\,\lambda$5721\,\AA & 0.29  $\pm$ 0.04 & 0.22\\
$[\ion{Fe}{vii}]\,\lambda$6087\,\AA & 0.44  $\pm$ 0.07 & 0.31\\
$[\ion{O}{i}]\,\lambda$6300\,\AA & 0.46  $\pm$ 0.02 & \hspace*{-0.15cm}0.3\\
$[\ion{Fe}{x}]\,\lambda$6375\,\AA & 0.07  $\pm$ 0.01 & 0.05\\
H$\alpha$ & \hspace*{0.15cm}4.7  $\pm$ 0.05 & 2.87\\
$[\ion{N}{ii}]\,\lambda$6583\,\AA & \hspace*{0.15cm}5.6  $\pm$ 0.06 & 3.41\\
$[\ion{S}{ii}]\,\lambda$6716\,\AA & 1.04  $\pm$ 0.05 &  0.62\\
$[\ion{S}{ii}]\,\lambda$6731\,\AA & 1.29  $\pm$ 0.07 & 0.77\\
\hline
\end{tabular}
\end{center}
\end{minipage}
\end{table}

\subsection{Reddening distribution}
\begin{figure}[h!]
\centering
 \resizebox{\hsize}{!}{\includegraphics[angle=-90]{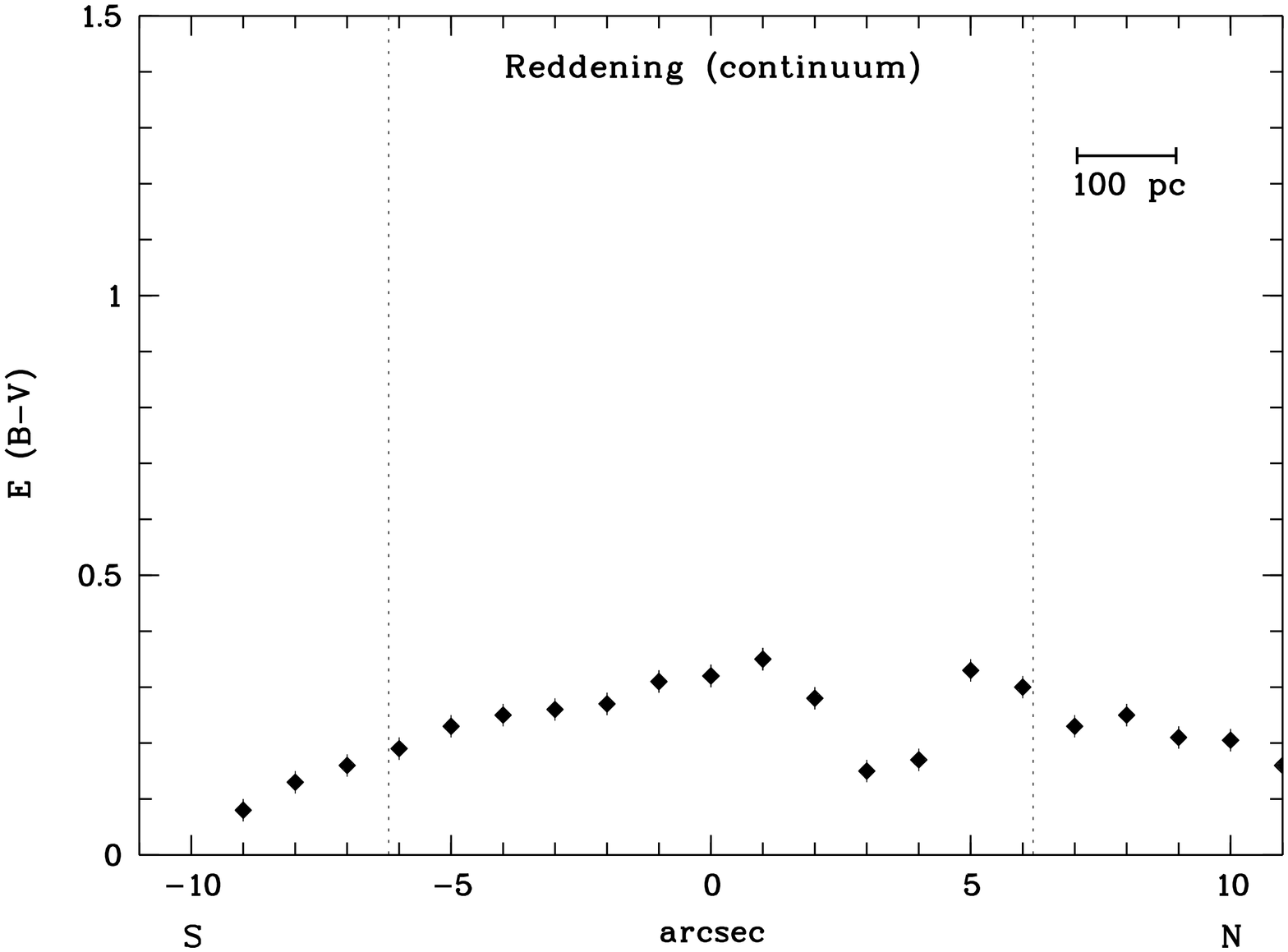}}
 \resizebox{\hsize}{!}{\includegraphics[angle=-90]{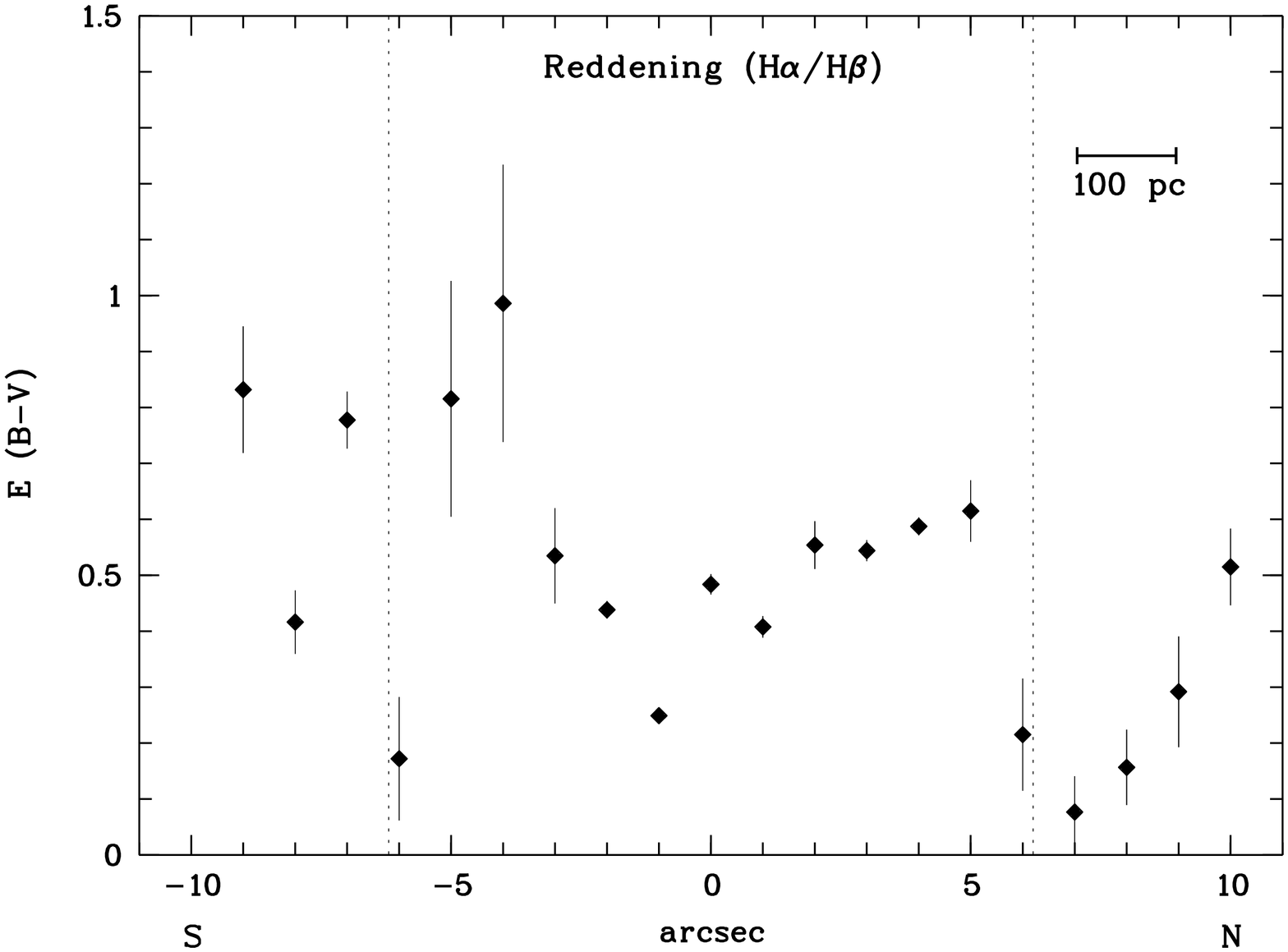}}
\caption{\label{slopereddening}
{\it Upper panel:} Reddening distribution of the continuum of the central
spectra  relative to that of the
stellar template. {\it Lower panel:} Reddening distribution derived from the
recombination value H$\alpha$/H$\beta$.
In both panels, the edge of the NLR as determined from the 
  diagnostic diagrams is indicated by dotted lines.}
\end{figure}

Two different measures of the reddening distribution were derived:
(i) the above mentioned reddening of the continuum slope in the
central parts of \object{NGC\,1386} with respect
to the template derived in the outer parts of the galaxy;
(ii) the reddening distribution obtained from the recombination
value for the intensity ratio H$\alpha$/H$\beta$.
Both reddening distributions are
shown in Fig.~\ref{slopereddening}. 

As already noted by \citet{wea91}, 
the continuum gets steadily redder towards
the nucleus. The reddest continuum  is found 1\arcsec~north of the
nucleus. We observe a significant drop 3--4\arcsec~north
of the nucleus not reported by \citet{wea91}. It
can be directly attributed to a ``blue'' continuum
seen 3\arcsec~north of the nucleus 
in the [F547M/F791W] color map by \citet{fer00}.
The blue continuum is associated with the northern tip of the central
emission--line structure in the [\ion{O}{iii}] image (Fig.~\ref{slit}).

As the match between the absorption lines of the stellar template and
those seen in the spectra is quite close, we believe that reddening by dust 
is the cause of the spatially varying continuum slope
and not an intrinsically redder stellar
population in the central part.

We cannot compare the absolute values of $E_{(B - V)}$ directly
as the reddening determined from the continuum
slope is a value relative to the reddening of the outer region
in the galaxy, i.e. the template. (For comparison, the smallest
reddening correction $E_{(B - V)}$ needed to fit the continuum of the stellar
template to that of the observed central spectra was set arbitrarily to zero in
Fig.~\ref{slopereddening}, {\it upper panel}.)
However, the reddening value obtained from the emission--line ratio
H$\alpha$/H$\beta$ does not show a comparable regular distribution.
Moreover, the reddening of the continuum slope covers a range of
$\sim$ 0.3 in $E_{(B - V)}$, while a three times larger range is covered
in the reddening distribution derived from the Balmer decrement ($\Delta E_{(B -
  V)}$ $\sim$ 1).  This can be due to
extinction by foreground dust in e.g.~the host galaxy which affects both the
template and the central spectra and thus do not reflect in the relative
reddening value determined from the difference in continuum slope of the
template and the NLR spectra.

It seems that the stellar population and the NLR are suffering different dust
extinctions. Possible explanations are
(i) high density leading to an additional collisional
contribution to the observed H$\alpha$/H$\beta$ ratio independent from
dust. However, extremely high densities are needed to explain the large
ratios and there is no observational evidence for such high densities.
(ii) Patchy dust clouds beyond the NLR may absorb the emission lines along
the line--of--sight of the NLR clouds. Also this explanation is rather unlikely.
We favor scenario (iii): The dust is intrinsic to the NLR
clouds and its column density 
varies along the
line of sight, resulting in the observed differences in
the reddening of the NLR and the stellar template.

We use the reddening distribution determined from the 
H$\alpha$/H$\beta$ emission--line ratio to correct for the intrinsic
reddening of the NLR itself 
as these lines originate in the NLR and thus give a better estimate for
the reddening within the NLR than the one determined from the continuum slope.
\subsection{Spatially resolved spectral diagnostics}
\label{2ddiag}
Diagnostic line--ratio diagrams
of the type pioneered by \citet{bal81} 
are commonly used to distinguish
bet\-ween emission--line object classes (e.g. Seyfert galaxies, LINERs, Starbursts,
transition objects), referring to a ``total'' spectrum
or the central spectrum of {\it one} object.
Here, we make use of the spatially resolved spectra
to discriminate
between different regions in the {\it same} object, for example the
NLR and circumnuclear or extended \ion{H}{ii} regions.
Thus, to probe the ``real'' NLR size, i.e. the central region which
is photoionised by the AGN, and to discriminate the contribution from
circumnuclear starbursts, we 
determined spatially resolved emission--line ratios. 

\begin{figure}[h!]
\centering
 \resizebox{\hsize}{!}{\includegraphics[angle=-90]{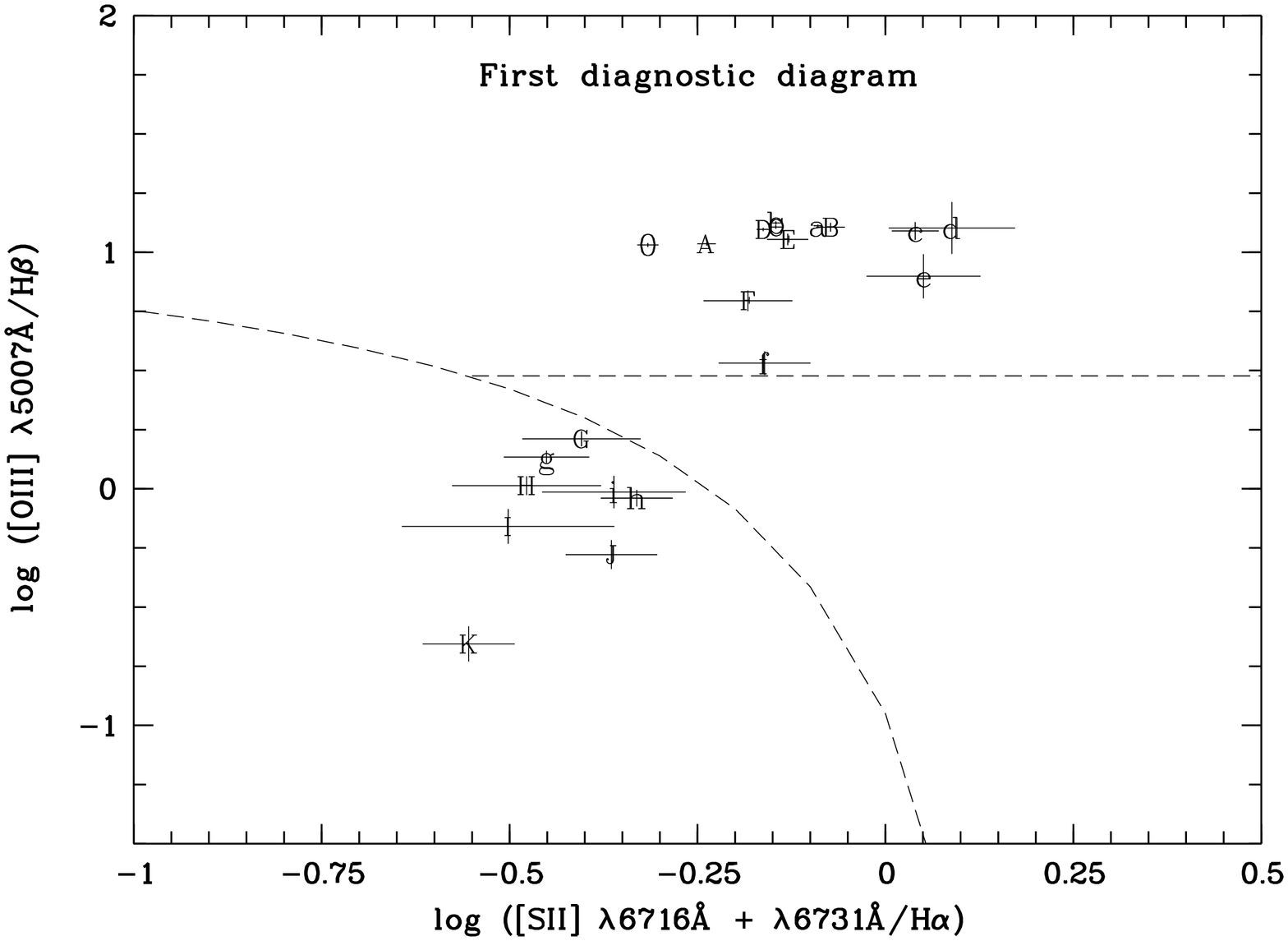}}
 \resizebox{\hsize}{!}{\includegraphics[angle=-90]{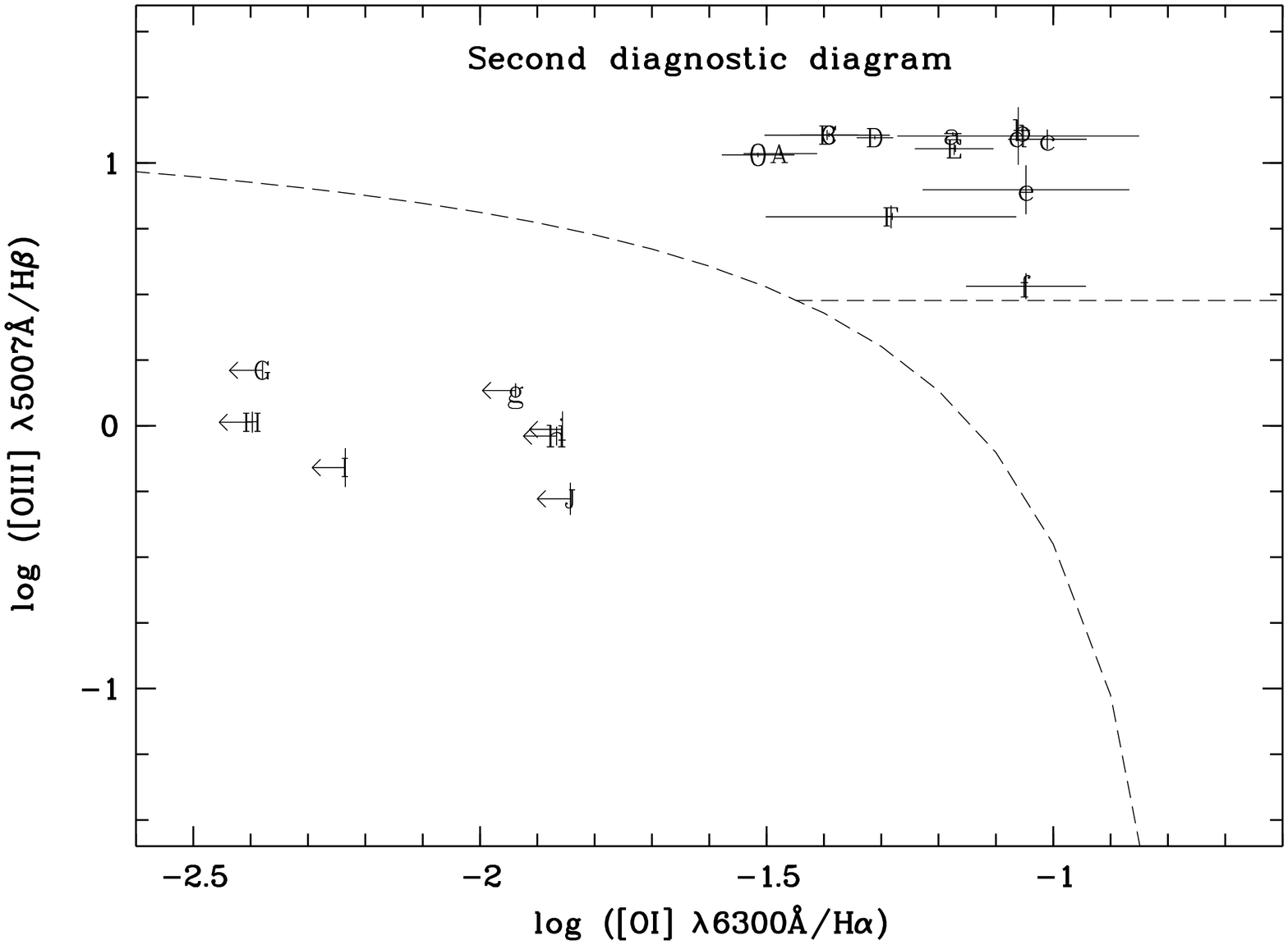}}
 \resizebox{\hsize}{!}{\includegraphics[angle=-90]{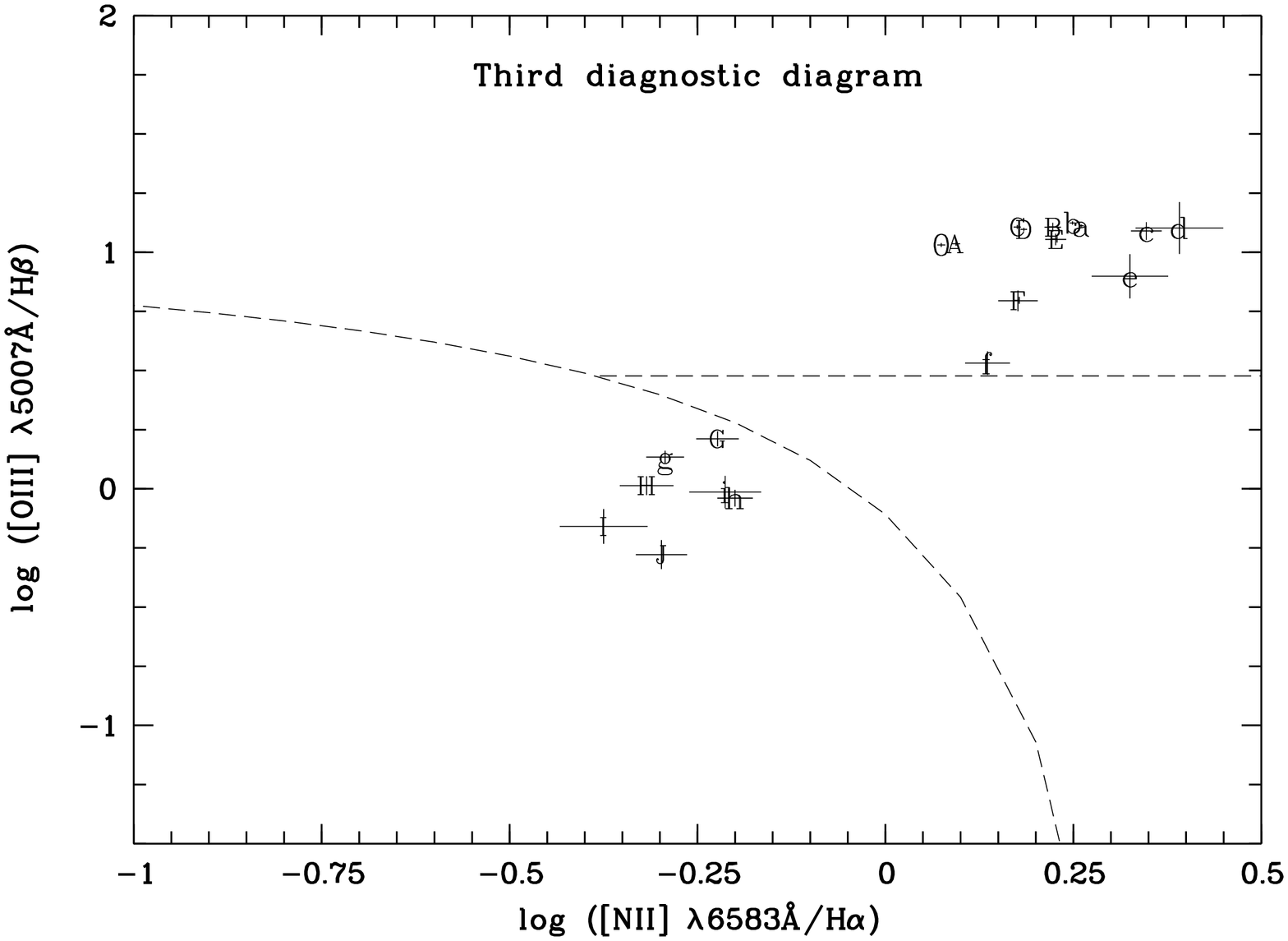}}
\caption{\label{diag} Diagnostic diagrams for
spatially resolved line ratios in \object{NGC\,1386}.
The dividing lines were taken from the analytic AGN diagnostics
of \citet{kew01}. The symbols are chosen such that ``0'' refers to
the central spectrum, the small letters mark southern regions, the capital
ones northern regions (``a,A'' = 1\arcsec~distance from the nucleus = 52\,pc, 
``b,B'' = 2\arcsec~distance = 104\,pc etc.). 
In all three diagrams, only the central $r \sim$ 6\arcsec~(310\,pc; A,a -- F,f)
show ratios expected for AGN--photoionised gas (upper right corner).
Further out (until $r \sim 10$\arcsec; G,g -- J,i), the line ratios fall
into the \ion{H}{ii}--region regime (lower left corner). 
}
\end{figure}

In Fig.~\ref{diag}, we present three diagnostic diagrams of
\object{NGC\,1386}. The high S/N ratio of our spectra enables us 
to measure line ratios for all three diagrams out to 10\arcsec~from the
nucleus. 
The symbols are chosen such that ``0'' refers to
the central spectrum, the small letters mark southern regions, the capital
ones northern regions (``a,A'' = 1\arcsec~distance from the nucleus = 52\,pc, ``b,B'' =
2\arcsec~distance = 104\,pc etc.).
In the second diagnostic diagram, the data
points of the outer 7--10\arcsec~are upper limits, due to the
faintness of the [\ion{O}{i}]\,$\lambda$6300\,\AA~line (hereafter  [\ion{O}{i}]) involved.
The line ratios of the central regions up to
6\arcsec~lie in the AGN regime (0, A/a -- F/f).
The outer line ratios extending to 10\arcsec~(G/g -- J/i)
all lie in the lower left corner usually 
covered by \ion{H}{ii} regions. This is a rather sharp transition
which can be interpreted as the edge of the AGN photoionised region,
i.e. the NLR, occuring at a radius of 6\arcsec~(310 pc), while emission
extending to 10\arcsec~originates from
circumnuclear \ion{H}{ii} regions. These \ion{H}{ii} regions can be attributed
to the \ion{H}{ii} regions seen by \citet{tsv95} in their
H$\alpha$+[\ion{N}{ii}] image, occuring at distances of
$\sim$6--11\arcsec~from
the nucleus along a p.a. of 5\degr~as used in our observations.

Our results agree with that of \citet{wea91} who
report an NLR extension of $\sim$6\arcsec~based on the third diagnostic
diagram. In the first and second diagnostic diagram, their data only include
the nuclear region due to the faintness of the involved
[\ion{O}{i}]
and [\ion{S}{ii}] lines in the outer regions.
In their third diagnostic diagram,
\citet{wea91} find some evidence for high [\ion{N}{ii}]/H$\alpha$ plus low
      [\ion{O}{iii}]/H$\beta$ ratios typical
for LINERs surrounding the highest ionisation gas near the nucleus. 
Our data do not show LINER--type ratios. The discrepancy between the
two results may be due to different stellar templates used. The influence of
the stellar template is discussed in Appendix~\ref{stellar}.

In Fig.~\ref{ratios}, we show the strongest reddening--corrected
line ratios as a function of radius.

Compared to pure imaging,
using diagnostic diagrams to determine the NLR size has the advantage to be
less sensitive to flux depth and to exclude contributions of circumnuclear starbursts.
Comparing the NLR size (radius 6\arcsec~$\simeq$ 310\,pc) determined from our diagnostic diagrams 
with literature values,  
we find that it is twice as large as the one found by \citet{sch03a}
(3\arcsec, Fig.~\ref{slit}), 
possibly due to the low sensitivity of the HST snapshot survey
(short integration time of 800\,s using the 2.4\,m HST
mirror compared to 1800\,s with the 8\,m VLT mirror and a 20 times larger
pixel size\footnote{Note, however, that the HST observations were an imaging campaign
  while we carried out spectroscopy, thus these values are not directly comparable.}).
\citet{fra03} detect line emission which they classify as
extended NLR out to a distance of
$\sim$10\arcsec~from the nucleus.
However, our analysis shows that the extended emission between 6\arcsec~and
10\arcsec~(310 -- 520\,pc) originates
from circumnuclear \ion{H}{ii} regions and cannot be attributed to the NLR.
Note that the total [\ion{O}{iii}] emission with a S/N $>$ 3 in our spectra
extends out to $r \sim$ 12\arcsec~, but 
only the central $r \sim$ 6\arcsec~can be attributed to the NLR.

As already pointed out  by \citet{wea91}, the sharp transition
can be interpreted as a real physical transition between
the AGN powered NLR and the surrounding \ion{H}{ii} regions in the disk.
\ion{H}{ii} regions may be
present over the entire emission--line region but 
inside a distance of 6\arcsec, the AGN ionisation dominates.
Therefore, it is reasonable to define the radius determined from the
transition seen in the diagnostic diagrams 
as the size of the NLR.
This conclusion is further supported by photoionisation models presented in
the following Section.
\begin{figure}[h!]
\centering
 \resizebox{\hsize}{!}{\includegraphics[angle=-90]{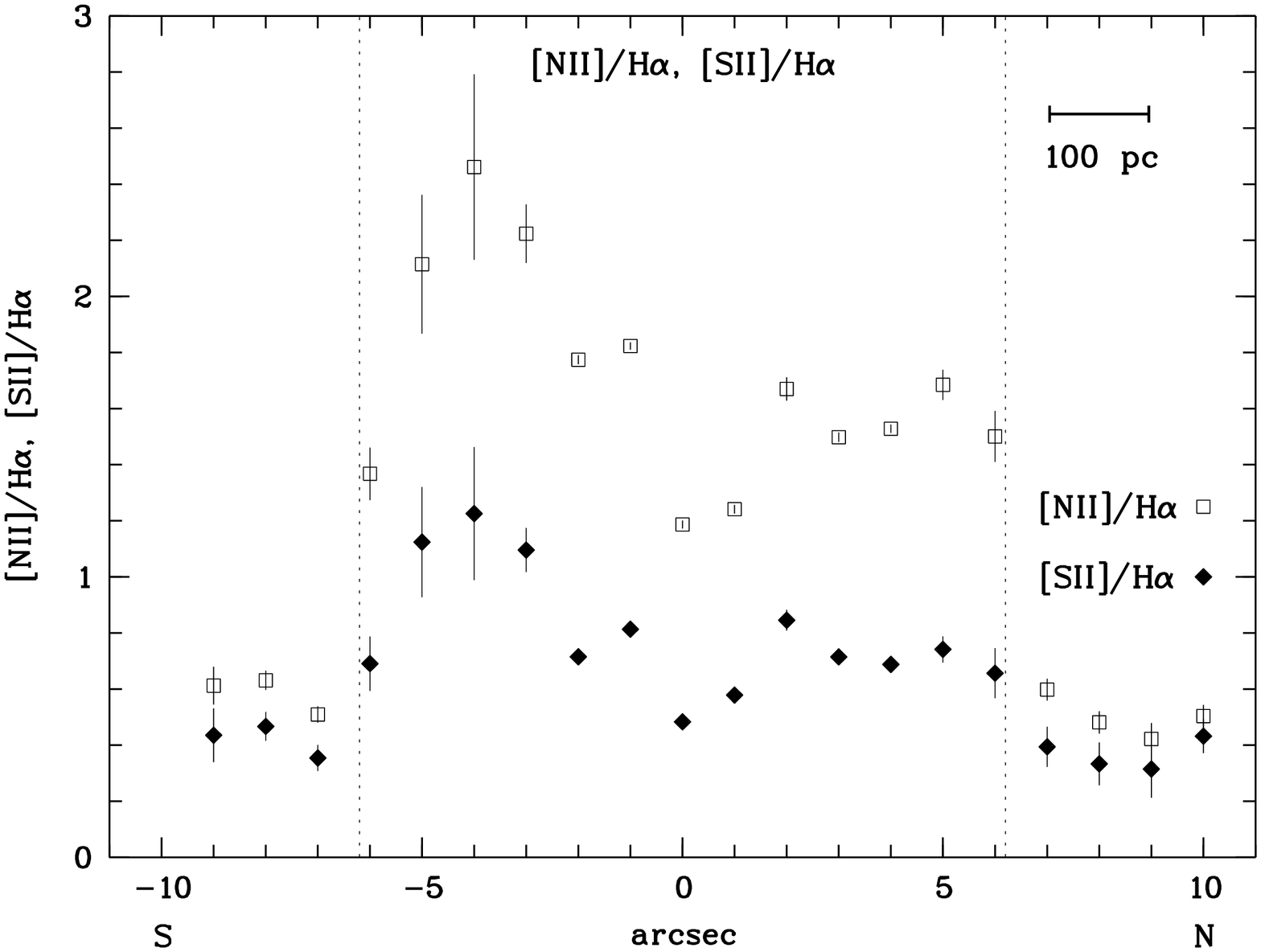}}
 \resizebox{\hsize}{!}{\includegraphics[angle=-90]{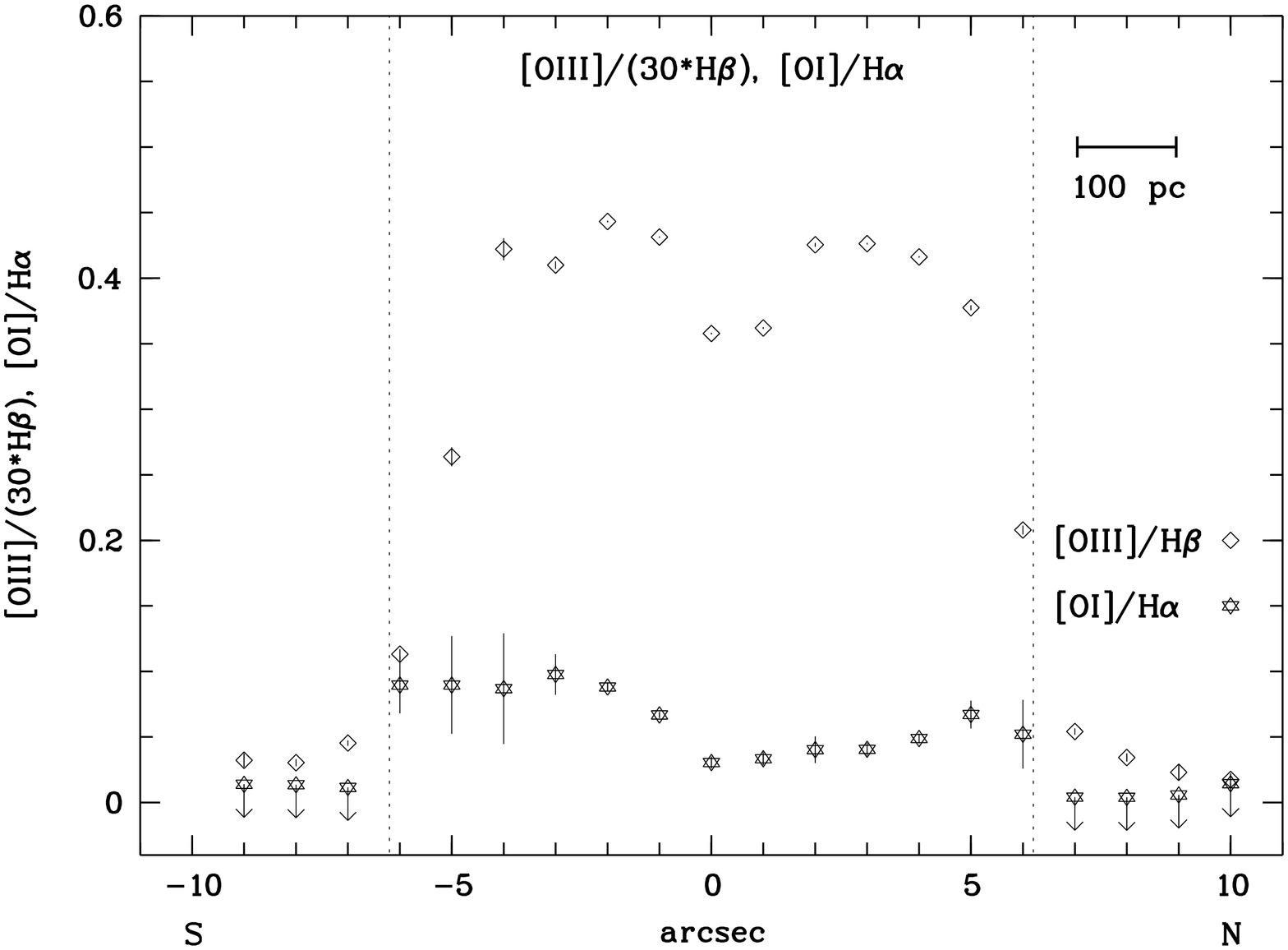}}
\caption{\label{ratios} Reddening corrected emission--line ratios used in the
  diagnostic diagrams as a function of distance
from the nucleus: [\ion{N}{ii}]/H$\alpha$ and [\ion{S}{ii}]/H$\alpha$ ({\it
upper panel}) as well as [\ion{O}{iii}]/H$\beta$ and [\ion{O}{i}]/H$\alpha$
 ({\it lower panel}). Note that the
[\ion{O}{iii}]/H$\beta$ ratio was divided by a factor of 30 for comparison.
The edge of the NLR as determined from the
  diagnostic diagrams is indicated by dotted lines.}
\end{figure}
\subsection{NLR size versus photoionisation modeling}
\label{cloudy}
The determination of the NLR size depends on the definition of the NLR itself.
We here assume that the NLR consists of gas photoionised
by the central AGN. Thus, we are able to determine the 
NLR size from the observed
transition of line ratios in the diagnostic diagrams from the AGN 
regime to that typical for \ion{H}{ii} regions (Fig.~\ref{diag}).

We have checked whether any mechanism exists
which drives line ratios in the diagnostic diagrams from the
AGN regime towards and into the regime usually covered by \ion{H}{ii} regions,
despite photoionisation by an intrinsic AGN continuum source.  
This question is of interest not only for the determination of
the NLR radius, but also for galaxy classifications based on diagnostic 
diagrams in general. 

Under the assumption of a single continuum source (the central AGN) being
responsible for the ionisation, 
the following parameters could potentially shift line ratios
from AGN-- towards \ion{H}{ii}--like in individual diagnostic diagrams: 
(i) extinction; (ii) ionisation parameter
in combination with (iii)
metallicity gradients; or (iv) electron density.  
We discuss each of them in turn. 

(i) Extinction can be excluded to have a large
effect on the observed line ratios in the diagnostic diagrams.
The line ratios in all three diagnostic diagrams
were chosen to minimize reddening effects by using neighbouring lines.
In addition, we measured the reddening distribution directly
(Fig.~\ref{slopereddening}, lower panel)
and use these results to correct for the reddening.
Using the different reddening distribution determined by the continuum slope
(Fig.~\ref{slopereddening}, upper panel) does not change the results within the errors.

\begin{figure}[h!]
\centering
 \resizebox{\hsize}{!}{\includegraphics[angle=-90]{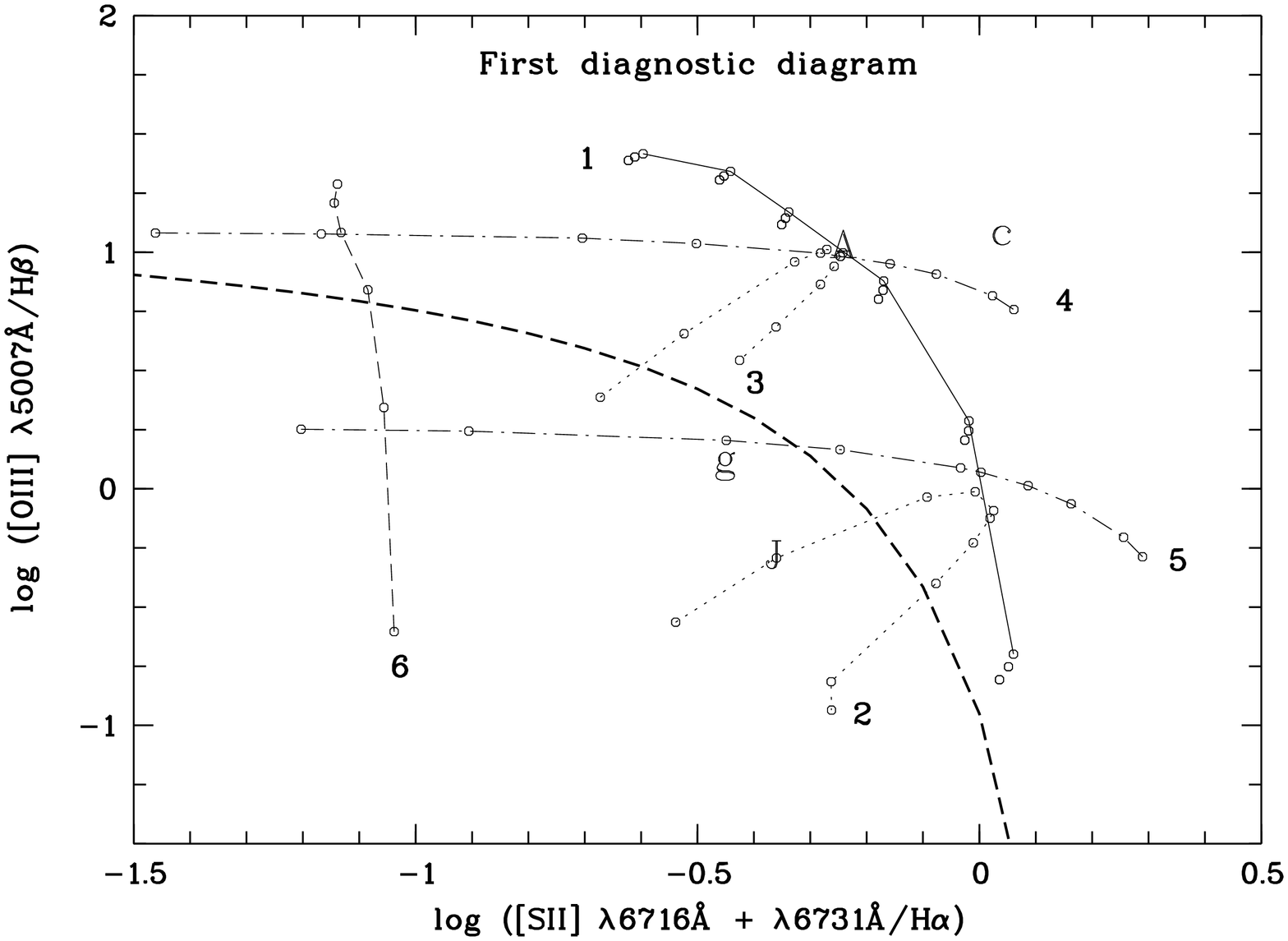}}
 \resizebox{\hsize}{!}{\includegraphics[angle=-90]{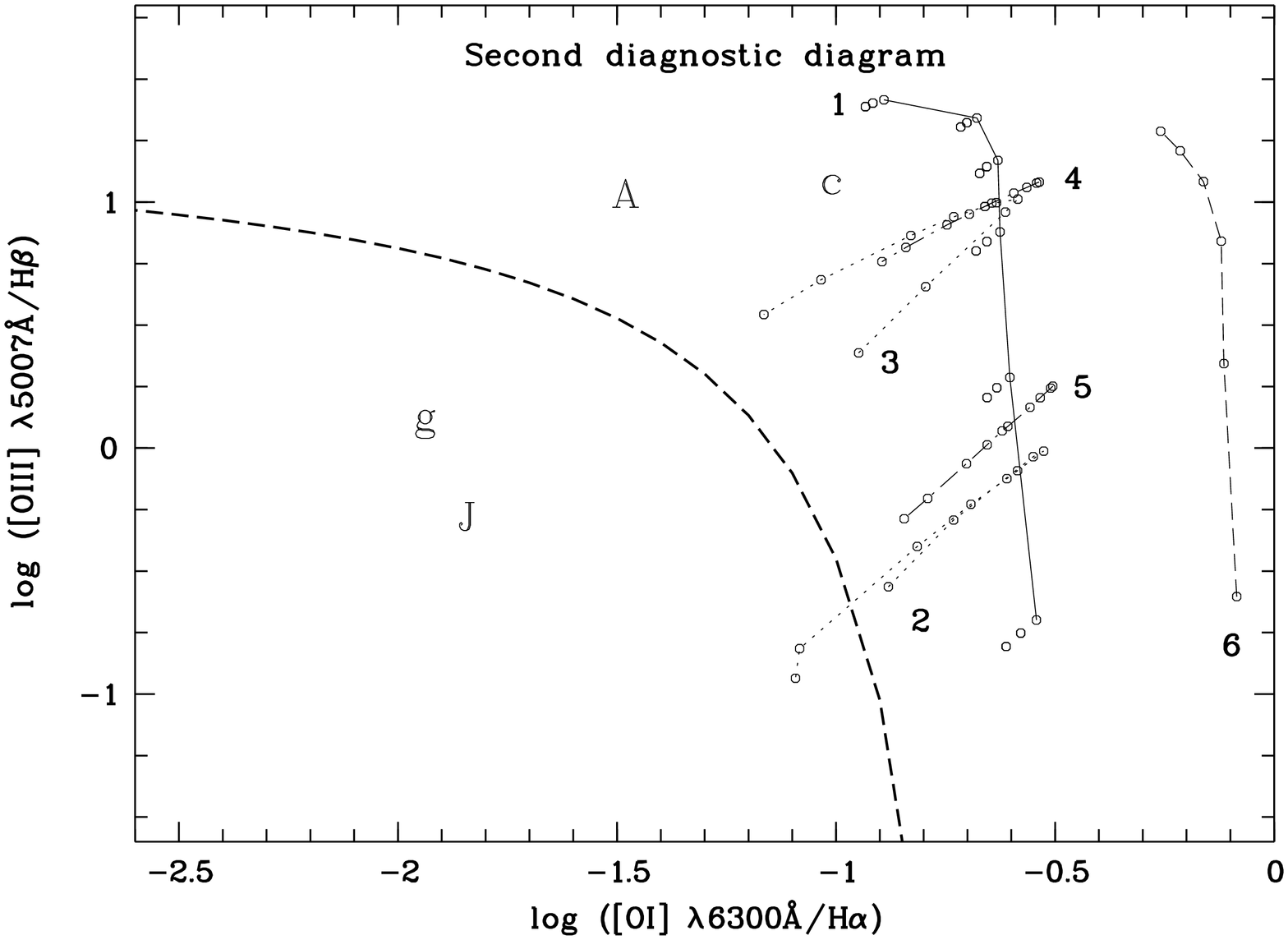}}
 \resizebox{\hsize}{!}{\includegraphics[angle=-90]{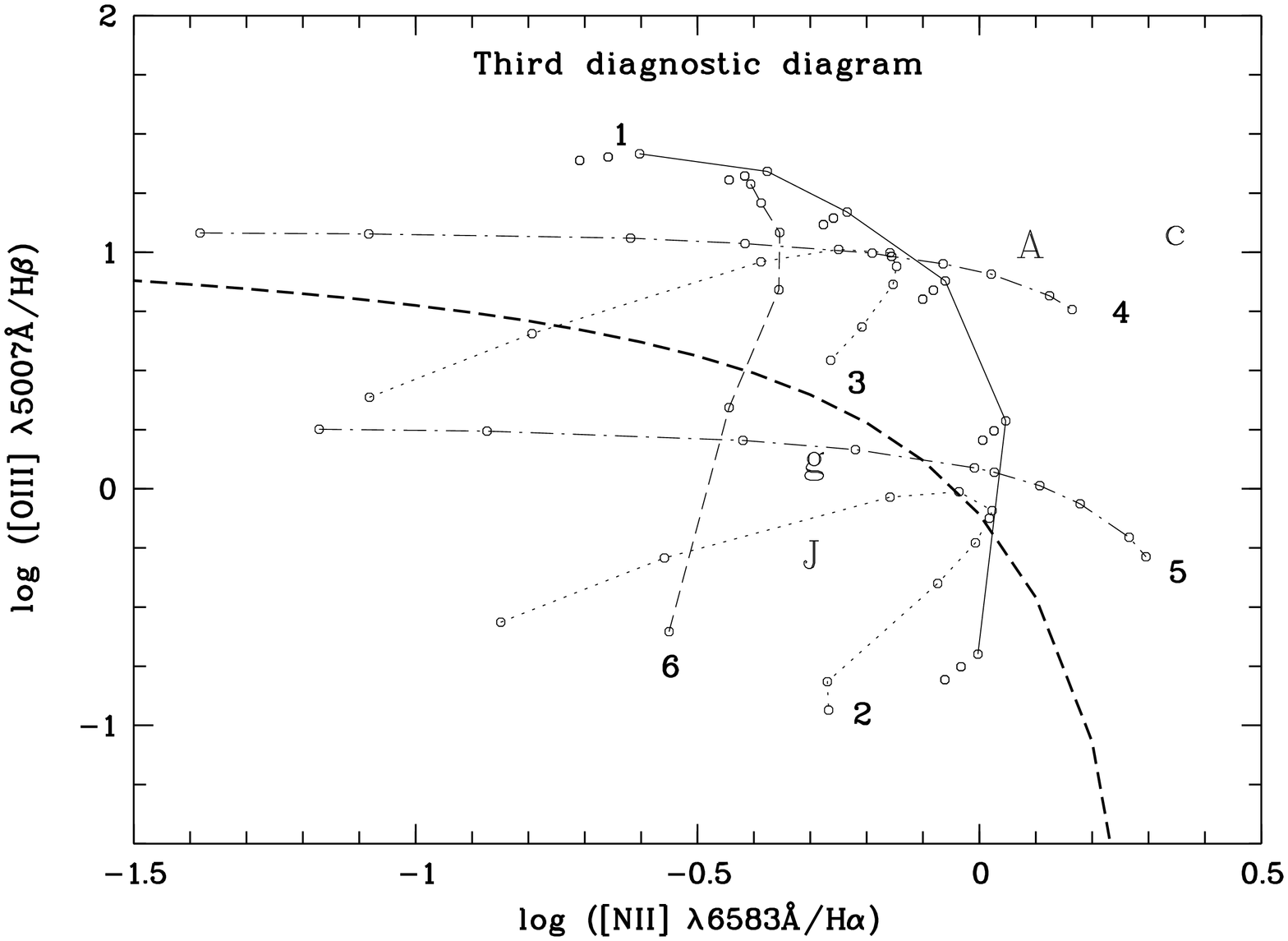}}
\caption{\label{cloudyfig} \texttt{CLOUDY} modeling results demonstrating the effects of
varying ionisation parameter (solid line 1), varying metal abundances 
(dotted lines 2 \& 3), varying N and S abundances (dash--dotted lines 4 \& 5),
and high density (dashed line 6) (see text for details). 
The bold dashed line indicates the separation between AGN regime
and \ion{H}{ii} regions. Four representative line ratios of
\object{NGC\,1386}  are also shown, falling in the AGN (A, c) and  \ion{H}{ii}--region regime (J, g).
}
\end{figure}

Models (ii)--(iv) presented below are based on the photoionisation code \texttt{CLOUDY}
[e.g. \citet{fer98}] and assume a pointlike AGN continuum source 
which illuminates clouds of constant density. The spectral--energy distribution
is a typical mean Seyfert continuum composed
of piecewise power laws \citep{kom97} with, in particular, an energy index
$\alpha$$_{uv-x}=-1.4$ in the EUV and a photon index $\Gamma$$_{x} =-1.9$
in X--rays. The clouds are assumed to be ionisation bounded and of solar
metallicity, unless stated otherwise. In Fig.~\ref{cloudyfig}, we present the
effects of varying ionisation parameter (solid line 1), varying metal
abundances (dotted lines 2 \& 3), varying nitrogen (N) and sulphur (S) abundances (dash--dotted
lines 4 \& 5), and high density (dashed line 6) on the emission--line ratios
in the diagnostic diagrams.

(ii) The ionisation parameter $U$ is defined as $U =
Q/(4 \pi c n_e r^2)$ with $Q =$ rate of H--ionising photons, $n_e$ =
electron density, and $r$ =
distance between photoionising source and emission--line clouds.
Variations in ionisation parameter alone have been studied by
many authors and generally lead into the LINER--region of diagnostic
diagrams for small ionisation parameters
under the assumption of solar abundances [e.g. \citet{sta84,ost89}]. This trend is shown in
Fig.~\ref{cloudyfig} (solid line 1) 
which includes the range of ionisation parameters we measure for \object{NGC\,1386},
varying from $\log U = -4.0 \ldots -1.5$ in steps of 0.5 (marked by circles) 
from bottom to top. Sequences were calculated three times, using 
densities $n$ = 800, 400, and 200\,cm$^{-3}$ 
and $r$ = 50, 100, and 250\,pc, respectively  
(printed as circles from right to left; 
only the data points for $n$ = 800\,cm$^{-3}$ and r = 50\,pc are connected by a line).
We come back to (ii) in combination with (iii$_{\rm b}$) below. 

(iii) The strongest hint argueing
against metallicity variations to change the line ratios
in the observed direction is the similarity of all three diagnostic diagrams.
In all three diagrams, the line ratios of
each spectrum fall in the same regime, either AGN or \ion{H}{ii} region. 
In fact, while extreme parameters may shift
line ratios towards \ion{H}{ii}--like in individual diagrams, no mechanism
does so for all three, as we will show in the following.

(iii$_{\rm a}$)  Metal abundances were varied between 
$Z$ = (0.05--3.0) $\times$ solar. Systematically decreasing abundances
first lead to an {\em increase} in line ratios which can be partly
traced back to oxygen being a strong coolant, and thus an increase of heating
in case of oxygen depletion (Fig.~\ref{cloudyfig}). 

The dotted lines 2 \& 3 show models of varying metal abundances
(3, 2.5, 1.7, 1.3, 1.0, 0.9, 0.5, 0.3, 0.1, 0.05 $\times$ solar; from right to left)
for constant ionisation parameter. The lower dotted line 2 corresponds
to $\log U = -2.8$, the upper dotted line 3 to $\log U = -3.7$.

(iii$_{\rm b}$) If, instead, only N and S abundances are varied,
line ratios shift horizontally in the [\ion{N}{ii}] and [\ion{S}{ii}] 
diagrams towards
the \ion{H}{ii} regime (Fig.~\ref{cloudyfig}).  
The dash--dotted lines 4 \& 5 show models of varying
N and S abundances 
(3, 2.5, 1.7, 1.3, 1.0, 0.9, 0.5, 0.3, 0.1, 0.05 $\times$ solar; from right 
to left) for constant ionisation parameter. The upper dash--dotted line 4 corresponds
to $\log U = -2.8$, the lower dash--dotted line 5 to $\log U = -3.7$.

Thus, a combination of outwards decreasing ionisation
parameter and decreasing metal abundances would place line ratios in
the \ion{H}{ii}--regime in the observed way, despite AGN--intrinsic excitation. 
However, the same trend does not 
hold for the [\ion{O}{i}] diagram. 
Just based on oxygen lines, this diagram is rather
insensitive to variations in N and S abundances and line ratios
remain well within the AGN regime for the whole parameter range 
(Fig.\ref{cloudyfig}). 
We conclude that combined ionisation parameter and metallicity effects cannot
explain the observations.   

(iv) Finally, the presence of a high density component will again shift
[\ion{N}{ii}] and [\ion{S}{ii}] towards the \ion{H}{ii} regime.
To demonstrate an extreme,
we plot in Fig.~\ref{cloudyfig} the case for $\log n_{\rm H}$ = 6 (dashed
line 6). The ionisation parameter varies from bottom to top in steps of 0.5 from
$\log U = -4.0 \ldots -1.5$. Again, while [\ion{N}{ii}]
and [\ion{S}{ii}] decrease,  [\ion{O}{i}] is strongly boosted, 
moving line ratios away from \ion{H}{ii}--like excitation. 
This effect can be traced back to the different critical densities.

We conclude that the observed distinction between \ion{H}{ii}--like and
AGN--like line ratios of \object{NGC\,1386} represents a true difference in
ionisation source, and thus our method to measure the NLR radius
is valid.  The second diagnostic diagram including 
the [\ion{O}{i}] emission--line,
new in our study compared to literature data, was essential to reach
this conclusion, since  
our photoionisation calculations show that 
there are means to reach the \ion{H}{ii}--regime with an AGN
as ionising source in the [\ion{N}{ii}] and [\ion{S}{ii}] diagrams.  

\subsection{Surface--brightness distribution}
\begin{figure}[h!]
\centering
 \resizebox{\hsize}{!}{\includegraphics[angle=-90]{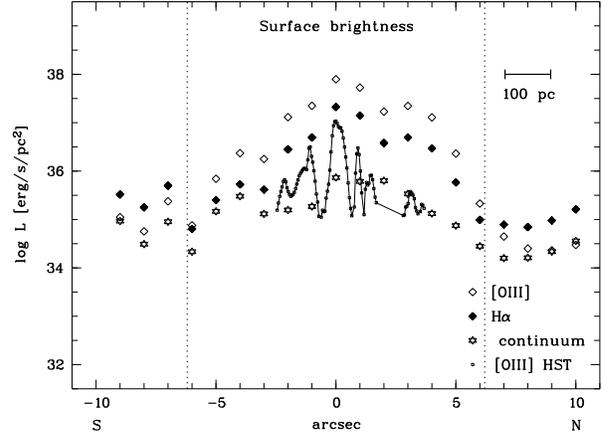}}
\caption{\label{lum} Surface--brightness distribution of \object{NGC\,1386} in
  [\ion{O}{iii}] (open diamonds), H$\alpha$ (filled diamonds), and continuum (at
  5450--5700\,\AA, stars). Error bars are smaller than the symbol size.
The [\ion{O}{iii}] surface--brightness distribution
  from the HST image with S/N $>$ 3 is shown as small open squares connected by a line.
The HST image has a 20 times higher spatial resolution but
a significantly lower sensitivity, not allowing to measure the outer parts
of the NLR. The edge of the NLR as determined from the
  diagnostic diagrams is indicated by dotted lines. 
}
\end{figure}

In Fig.~\ref{lum}, we present the surface--brightness distribution
of the [\ion{O}{iii}] and H$\alpha$ emission line as well as of the continuum (at
  5450--5700\,\AA).
The surface--brightness distributions are similar to each other, centrally peaked and 
all decreasing with distance from the
nucleus. They reveal a secondary peak $\sim$3\arcsec~north of the nucleus and a
slightly lower tertiary peak $\sim$4\arcsec~south of the nucleus.
While the [\ion{O}{iii}] surface brightness is exceeding that of H$\alpha$ in
the central parts, this behaviour changes
at the edge of the NLR as determined from our 2D
diagnostic diagrams (indicated by the dotted lines).

For comparison, the [\ion{O}{iii}] surface--brightness distribution
from the HST image of \citet{sch03a} is also shown. 
It was derived by averaging three vectorplots along the major axis of the NLR
emission along p.a. = 5\degr$\pm$1\degr. It clearly shows the
higher spatial resolution of the HST image (0\farcs0455
pix$^{-1}$) compared to the 1\arcsec~spatial sampling of our spectral data.
However, as already mentioned in Section~\ref{2ddiag}, 
it once again reveals the low sensitivity of the HST image
compared to our spectroscopy as it detects 
[\ion{O}{iii}] emission at a S/N of 3 only out to 3--4\arcsec~from the nucleus.
The HST [\ion{O}{iii}] surface--brightness distribution reveals several
subpeaks of possibly individual NLR clouds, as can be already seen in the
[\ion{O}{iii}] image (Fig.~\ref{slit}).
These substructures are smoothed out in the $\sim$20 times lower spatial resolution
of our spectra.
\subsection{Electron--density distribution}
\begin{figure}[h!]
\centering
 \resizebox{\hsize}{!}{\includegraphics[angle=-90]{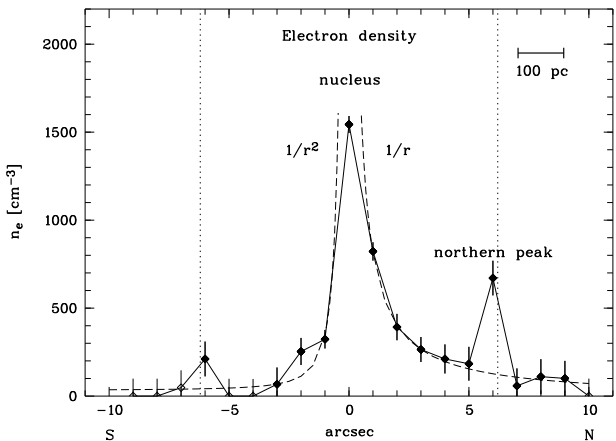}}
\caption{\label{density} Electron density obtained
from the [\ion{S}{ii}]\,$\lambda$6716\,\AA/$\lambda$6731\,\AA~ratio as a function of distance from the nucleus.
Open symbols indicate locations
where $n_e$ is in the low--density limit (assumed $\le$50\,cm$^{-3}$).
The electron density decreases with radius and shows a secondary peak 
6\arcsec~north of the nucleus as well as a tertiary fainter one at
6\arcsec~south. 
The fits $n_e (r) = n_{e,0}
  r^{-1}$ and $n_e (r) = n_{e,0}
  r^{-2}$ are shown as dashed lines. The edge of the NLR as determined from the
  diagnostic diagrams is indicated by dotted lines.}
\end{figure}

Applying the classical methods outlined in \citet{ost89},
we derive the electron density as a function of distance
from the nucleus using the ratio of the 
[\ion{S}{ii}]\,$\lambda$$\lambda$6716,6731\,\AA~(hereafter  [\ion{S}{ii}])
pair of emission lines (Fig.~\ref{density}).
We used the observed central temperature to correct for the dependency of electron
density on temperature\footnote{$n_e ({\rm T}) 
= n_e ({\rm obs}) \cdot \sqrt{(T/10000)}$} ($T_e$ $\simeq$ 15650
K). Due to the faintness of the involved
[\ion{O}{iii}]\,$\lambda$4363\,\AA~emission line, we were not able to measure
the temperature in the outer parts.

The electron density is highest at the nucleus with 
$n_e$ $\simeq$ 1540\,cm$^{-3}$ and decreases outwards down to the low--density limit
(assumed to be 50\,cm$^{-3}$). A secondary peak can be seen at a distance of
6\arcsec~to the north of the nucleus ($n_e$ $\simeq$ 670\,cm$^{-3}$).\footnote{
Note that the temperature can be a function of distance from the central
AGN and is most probably decreasing, if the AGN is the only heating source.
In such a case, correcting with the central temperature overestimates
the electron density in the outer parts. Thus, a decreasing temperature can steepen the slope
of the electron density.}

Comparing the position of the observed northern $n_e$ peak with the 2D
electron--density 
distribution of \citet{wea91}, the closest data point
is at $\sim$6\farcs5~north--east from the center at roughly a p.a. of
3\degr~with $n_e \le 100$\,cm$^{-3}$ (their Fig. 10d). 
The spatial distance between this data point and our $n_e$ peak
is $\sim$25 pc. Thus, the $n_e$ peak observed in our spectra may be
attributed to an individual NLR cloud or local density inhomogeneities which
our line--of--sight happens to pass.
\citet{mau92} also detect individual NLR clouds using speckle interferometry
and conclude that the ionising radiation must be absorbed on scales $<$15 pc
in a clumpy structure of the NLR.
Interestingly, the peak occurs at the edge of the NLR. 
We discuss the origin of the northern peak in greater detail in
Section~\ref{northern}. 
Another, but significantly smaller peak is visible at the southern edge
($n_e$ $\simeq$ 210\,cm$^{-3}$).

While a detailed fit of
a power--law function is complicated by  the intrinsic
scatter of the data points, a general dependency on
$r$ can be estimated. 
The best fits of a power--law function with
$n_e (r) = n_{e,0} r^{-\delta}$ were derived with $\delta$ between $1$ and $2$
for the electron density (Fig.~\ref{density}),
neglecting the data point 6\arcsec~north of the nucleus which
may have another origin (Section~\ref{northern}).

\subsection{Ionisation--parameter distribution}
\begin{figure}[h!]
\centering
 \resizebox{\hsize}{!}{\includegraphics[angle=-90]{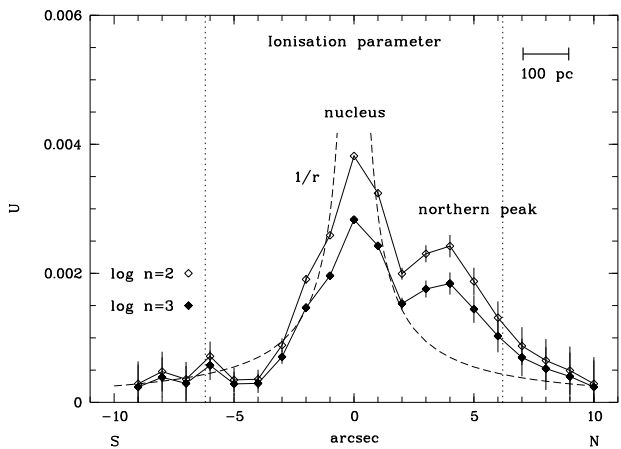}}
\caption{\label{ioni}Ionisation parameter
derived from [\ion{O}{ii}]/[\ion{O}{iii}] ratio
as a function of distance from the nucleus (open symbols: $n_H$ =
100\,cm$^{-3}$, filled ones: $n_H$ = 1000\,cm$^{-3}$).
The ionisation parameter is highest in the center
and decreases with distance with a secondary peak at 4\arcsec~to the north of the
center.  The fit $U(r) = U_{0}
  r^{-1}$ is shown as dashed line. The edge of the NLR as determined from the
  diagnostic diagrams is indicated by dotted lines.
}
\end{figure}

To estimate the value of the ionisation parameter, the line ratio
[\ion{O}{ii}]$\lambda$3727\,\AA/[\ion{O}{iii}]\,$\lambda$5007\,\AA~(hereafter
 [\ion{O}{ii}]/[\ion{O}{iii}]) can be used
[e.g. \citet{pen90}]. \citet{kom97} have shown that
in case of low density
($n_H < 10^3$\,cm$^{-3}$), this line ratio yields $U$
independent of the shape of the ionising continuum.
It gives a good
approximation of the ionisation parameter as long as there are no strong
density inhomogeneities with radius. 
If the NLR contains a mixture of densities, the ratio predicts
a too high value of $U$. But even if the absolute value of $U$ was
overpredicted, the observed slope would not change.
We used the theoretical variations 
of [\ion{O}{ii}]/[\ion{O}{iii}] calculated by \citet{kom97} (their Fig. 7)
with $U$ for a power--law continuum ($\alpha_{\rm uv-x}$ = $-$1.5) and
a density of $2 < \log n_H$ (cm$^{-3}$) $< 3$ (Fig.~\ref{ioni}). 
As the electron density is  decreasing with distance from
the nucleus ($\sim$1500 to $\sim$50\,cm$^{-3}$, Fig.~\ref{density}), the resulting
ionisation parameter lies in between the two curves calculated for $\log
n_H = 2$ and  $\log n_H = 3$, i.e. the slope narrows.

While \citet{wea91} propose a decreasing ionisation parameter to explain the
observed line ratios in \object{NGC\,1386}, we can determine the ionisation parameter directly and
indeed find that it is decreasing with distance from the
nucleus. It varies between $U_{\rm log (n_e) = 3}$ = (2.83$\pm$ 0.01) $\cdot 10^{-3}$ and $2.4 \cdot 10^{-4}$ within the NLR
with the highest value in the center.
It reveals a secondary peak
at 4\arcsec~to the north of the center
which is at the north edge of a ``blob'' of emission clearly
visible in the [\ion{O}{iii}] image from \citet{sch03a} (Fig.~\ref{slit}).
Compared to the electron density, the secondary peak is 2\arcsec~closer to the
nucleus and significantly broader. We discuss the origin in Section~\ref{northern}.
In Appendix~\ref{ionired}, we discuss any possible influence of the reddening
distribution on the slope of the ionisation parameter.

We estimated the general dependency of the decrease of $U$ on
$r$. The best fit of a power--law function with
$U(r) = U_{0} r^{-\delta}$ is derived with $\delta \sim 1$
(Fig.~\ref{ioni}), again neglecting the ``northern peak''.
Inserting the fit  $n_e \propto r^{-1}$ 
to the definition of $U = Q/ (4 \pi c n_e r^2)$ yields
$U \propto r^{-1}$. This is in agreement with
the fit to the ionisation parameter. It indicates that the excitation of the
bulk of the
NLR is due to photoionisation and that ionisation by e.g. shocks is negligible.
\subsection{Northern peak: evidence for shocks}
\label{northern}
Our observations show that both the electron density and
the ionisation parameter are decreasing with radius.
They show a secondary peak $\sim$ 6\arcsec~and 4\arcsec~north of
the center, respectively. 
The peak at 4\arcsec~is close to the northern tip of the 
[\ion{O}{iii}] emission (Fig.~\ref{slit}).
In the same region, a blue continuum indicating low extinction is observed  
[Fig.~\ref{slopereddening}, {\it upper panel}, see also \citet{fer00}].

Can both the increased density and the ionisation parameter
originate from a shock of a radio
jet? Extended radio emission has been found by \citet{nag99},
displaced to the observed slit p.a. tracing
the major [\ion{O}{iii}] extension by $\sim$15\degr~(p.a.$_{\rm radio}$ =
170\degr~versus p.a.$_{\rm [OIII]}$ = 5\degr).
A scenario in which
the outflowing material causing the radio emission
compresses the line--emitting gas by shocks,
increasing the ambient density, seems
to be rather common for Seyfert galaxies
[e.g. \citet{cap96,fal98}].
While the shock front
results in a locally increased
density, the continuum of the shock can 
possibly ionise a larger area of the surrounding medium 
and produce a change in the observed line ratios,
resulting in an $U$ bump which is broader than the sharp peak of $n_e$.
Fast autoionising shocks can produce such a local continuum ionising the
surrounding medium \citep{dop96}.

To check the shock scenario, 
we searched for high--ionisation
emission lines (e.g. [\ion{Fe}{x}]\,$\lambda$6375\,\AA). 
In the nuclear spectra, we find 
strong [\ion{Fe}{vii}]\,$\lambda$5721\,\AA~and  [\ion{Fe}{vii}]\,$\lambda$6087\,\AA~emission, comparable to that of [\ion{O}{i}]. 
Due to our low spectral resolution, [\ion{Fe}{x}]\,$\lambda$6375\,\AA~is blended with the [\ion{O}{i}]\,$\lambda$6363\,\AA~line. 
Thus, to estimate the contribution of 
[\ion{Fe}{x}]\,$\lambda$6375\,\AA, we fit two Gaussians
to [\ion{O}{i}]\,$\lambda$$\lambda$6300,6363\,\AA~with line fluxes fixed to the ratio of 1:3
and a third one to fit any remaining flux of [\ion{Fe}{x}]\,$\lambda$6375\,\AA.
Indeed, we find evidence for [\ion{Fe}{x}] contribution in 
the central 2\arcsec~([\ion{Fe}{x}]/H$\beta$ $\sim$ 0.047 in the center,
[\ion{Fe}{x}]/H$\beta$ $\sim$ 0.049 at 1\arcsec~to the
north) and at 6\arcsec~([\ion{Fe}{x}]/H$\beta$ $\sim$ 0.049), i.e. at the same region where
we observe an increased electron density.

The observed minimum in the reddening of the continuum slope 
roughly 3--4\arcsec~north
of the nucleus (Fig.~\ref{slopereddening}, {\it upper panel})
could be mimicked by an additional local continuum contribution.
The extinction in the NLR 
measured from the H$\alpha$/H$\beta$ value reaches a minimum at 
6--7\arcsec~north of the nucleus (Fig.~\ref{slopereddening}, {\it lower
  panel}). This can be explained by the destruction of dust by local shocks.

To conclude, we find several signs for shocks.
A radio jet interacting with the NLR gas is a plausible explanation for
the northern peak observed in the electron density and the ionisation--parameter distribution.
\subsection{NLR versus stellar velocity curve}
We derived the NLR velocity curve from the average of the peak wavelengths of
the H$\alpha$ and [\ion{N}{ii}] emission lines.
In addition, given the high S/N ratio of our spectra, we are able
to trace the velocity field from the ``peak wavelengths''
of the stellar absorption line \ion{Ca}{ii} K
(before subtraction of the stellar template)
throughout the whole region as this line is not
blended with emission lines.
The uncertainty in velocity values obtained by comparing repeated measurements
at the same point is 20\,km\,s$^{-1}$.
Note that gas velocities derived from the emission of [\ion{O}{ii}], H$\beta$, and [\ion{O}{iii}]
reveal a curve similar to that of H$\alpha$, [\ion{N}{ii}].

\begin{figure}[h!]
\centering
 \resizebox{\hsize}{!}{\includegraphics[angle=-90]{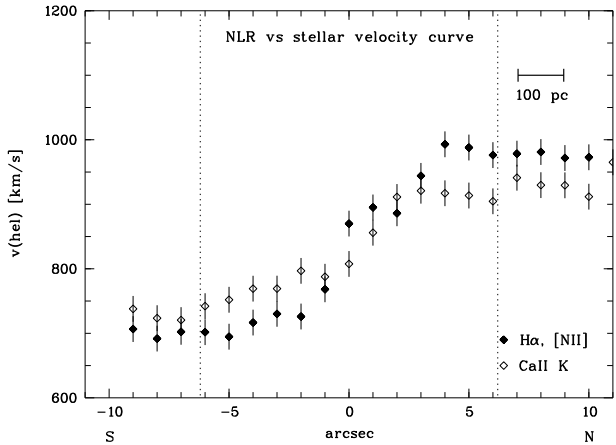}}
\caption{\label{velocity} Velocity curve of
the NLR derived from the average value of the peak wavelengths
of the H$\alpha$ and [\ion{N}{ii}] emission lines (filled symbols) as well as the stellar 
velocity curve measured using the \ion{Ca}{ii} K absorption--line 
``peak'' wavelength as seen in the ``raw'' spectrum (open symbols).
 The edge of the NLR as determined from the
  diagnostic diagrams is indicated by dotted lines.
}
\end{figure}

Both, the stellar and the gaseous velocity curves, have a
similar slope (Fig.~\ref{velocity}). 
The stellar velocity curve covers an overall smaller 
range of ($\sim$720--940)$\pm$20 km\,s$^{-1}$, while
the emission lines show velocities ranging from ($\sim$690--990)$\pm$20\,km\,s$^{-1}$. We define the kinematical center by taking the
symmetry center of the outer portions of the velocity curve.
The kinematical center has an heliocentric velocity of
$\sim825$$\pm$20\,km\,s$^{-1}$, 
i.e. slightly lower than the one given in
NED ($v_{\rm hel} = 868$\,km\,s$^{-1}$).
However, the velocity measured at the position of the optical nucleus is with $v_{\rm hel} =
870$$\pm$20\,km\,s$^{-1}$ the same as the literature value within the errors.

The kinematical center of the stellar velocity curve lies 
$\sim$1\arcsec~north of the optical nucleus (``0'' on the spatial scale).
This offset is in agreement with the analysis of the H$\alpha$ peak velocity field by
\citet{wea91} and \citet{sch03} who report an offset of the kinematical center
of $\sim$1\arcsec~to the north--east of the photometrical center.
The kinematical center is suggestive of a hidden nucleus
residing 1\arcsec~north of the optical nucleus
as has also been suggested by \citet{ros00} (see Fig.~\ref{slit}).

However, the kinematical center of the
NLR velocity curve we find at a
p.a. of 5\degr~[compared to 23\degr~of \citet{sch03}]
is $\sim$0\farcs5 south of the optical
nucleus. It may be indicative of 
a disturbed velocity field from outflows and shocks at our observed slit orientation.

While the kinematics of the emission--line region in \object{NGC\,1386} has been
investigated by several authors, an interpretation of the observed velocity
field is not unambiguous. 
Though the natural explanation is that of a slightly projected rotation curve
from an inclined emission--line disk in the center, an interpretation by a
bipolar flow accidentally aligned with the kinematical major axis of the
rotating disk is also possible \citep{sch03}. 
However, as the velocity curves of both NLR gas and stars are similar, a
bipolar outflow of the NLR is rather unlikely as it had to be decoupled from the stellar
gravitational field and its accordingly stratified
interstellar medium.

Instead, the NLR seems to follow the
stellar rotation indicating that the NLR gas is distributed in a
disk rather than a sphere. Galactic rotation of narrow--line gas seems not to be
uncommon [e.g. \citet{sch03}]. 
\citet{mul96b} propose that the NLR gas is distributed
in the galactic disk rather than a sphere, 
to explain differences of the observed NLR structure with that expected from the
Unified Model in the case of a spherical distribution of the NLR gas.

The gaseous velocity curve does not show a different behaviour
in the NLR and the \ion{H}{ii}--region regime. This indicates that the NLR does not
consist of an intrinsically different gas component with its own
  kinematics, but is ambient gas
in the galactic disk photoionised by the AGN.

Our results reveal some differences between the stellar and the
NLR velocities in \object{NGC\,1386}. 
The NLR velocities are in general higher than those of the
stellar absorption lines. The difference is especially large in both the  
outer southern and northern part of the NLR.
These can be due to 
many effects, e.g. rotating gas disk/ring within a kinematically hotter 
stellar component (the interpretation being 
complicated by the fact that due to the high inclination of the galaxy, 
the line of sight towards NLR intersects both the stellar disk and 
the bulge), gas moving along elliptical streamlines in a barred
potential (keep in mind the ambiguous Hubble--type classification, Sect.~\ref{literature}),
or outflow/jet interaction.
However, for a detailed interpretation of the stellar and gaseous velocity field,
kinematical modelling is required which lies beyond the scope of this paper.
\subsection{NLR size--luminosity relation}
The NLR size and its relation to [\ion{O}{iii}] luminosity
has been studied by \citet{ben02} and \citet{sch03b}
based on [\ion{O}{iii}] images. While \citet{ben02} find a relation between
NLR size and luminosity with a slope of 0.5,
\citet{sch03b} report a slope of 0.33.
If the NLRs were ambient circumnuclear gas fully photoionised
by the nuclear continuum source, a Str\"omgren--type radius--luminosity
relationship $R \propto L^{0.33} n^{-0.66}$ and thus a slope of
0.33 is indeed expected in the case of a constant density.
However, a decreasing density as observed for \object{NGC\,1386} seems to be common
to several Seyfert galaxies [e.g. \citet{sto92,fra00,fra03}],
contradicting the simple
``Str\"omgren explanation''.

A slope of 0.5 found by \citet{ben02} 
for the NLR size--luminosity
relation of both the [\ion{O}{iii}] emission line as well as H$\beta$ 
is expected if on average, all AGNs have
the same ionisation parameter, density, and ionising spectral energy
distribution \citep{net90, net04}.
It suggests a self--regulating mechanism that determines 
the size of the NLR to scale rather with the
ionisation parameter $U$:  $U \propto Q/( R^2 n)$ 
with $Q \propto L({\rm H}\beta)$ which leads to
$R \propto L({\rm H}\beta)^{1/2}$, for a given $U$ and $n_e$, in agreement 
with the empirical relation (assuming that $L({\rm H}\beta) \propto L(\rm [OIII]))$.
At the outskirts of the NLR, $n_e \sim 10^{2-3}$\,cm$^{-3}$ and $U \sim
10^{-(2-3)}$, so that efficient [\ion{O}{iii}] emission comes
from regions with $U n_e \sim 1$.
The NLR size--luminosity relation indicates that all
objects have a similar $U n_e$ value at the edge of
their NLR, independent of the individual spatial
behaviour of $U$ and $n_e$. Thus,
the observed spatially decreasing ionisation parameter
and density in \object{NGC\,1386} do not contradict the explanation
of the NLR size--luminosity relation in terms of
a constant (in all objects) ionisation parameter. 

However, the NLR size determination using [\ion{O}{iii}] images alone
bears various uncertainties due to e.g. the sensitivity dependency
of the observation and possible contributions from circumnuclear starbursts
as has been shown for \object{NGC\,1386}. Compared to the spatially resolved
spectral diagnostics measuring the ``real'' NLR size, the apparent NLR size determined
by [\ion{O}{iii}] images can be either smaller in case of low sensitivity 
or larger in case of contributions of circumnuclear starbursts. 
Using the methods described here to
determine the NLR sizes of 
a larger sample of objects will help
to scrutinise the NLR size--luminosity relation.
Nevertheless, individually differing NLR sizes may not change
the overall slope of the size--luminosity relation  which
extends over four orders of magnitudes in luminosity
and two orders of magnitudes in size.

\section{Conclusions}
Applying diagnostic diagrams to our spatially resolved spectra, 
we observe a transition of emission--line ratios from the central AGN region
to the surrounding \ion{H}{ii} regions. This transition occurs at $r \sim$ 
6\arcsec~(310 pc) to the
north and south of the optical nucleus and is, for the first time, observed in
all three diagnostic diagrams, i.e. including the second diagnostic diagram
involving the [\ion{O}{i}] emission line.
The most probable explanation for 
this transition is that the stellar ionisation field
dominates that of the AGN at a distance of $\ge$6\arcsec. We can exclude other
effects such as the variation of physical parameters from 
\texttt{CLOUDY} photoionisation modelling. The second diagnostic diagram
([\ion{O}{iii}]/H$\beta$ versus [\ion{O}{i}]/H$\alpha$) was essential
to draw this conclusion as our photoionisation calculations show that 
there are means to reach the \ion{H}{ii}--region regime with an AGN
as ionising source in the [\ion{N}{ii}] and [\ion{S}{ii}] diagrams.
We determine the radius of the NLR in \object{NGC\,1386} to 6\arcsec~(310\,pc), 
independent of sensitivity and excluding
[\ion{O}{iii}] contamination from circumnuclear starbursts.
Applying this method to a larger sample of Seyfert galaxies
will help to scrutinise the NLR size--luminosity relation.

We derive physical parameters of the NLR in \object{NGC\,1386} such as reddening,
surface brightness, electron density, and ionisation parameter as
a function of distance from the nucleus.
The differences between the reddening distributions 
determined from the continuum slope and the Balmer
decrement argue in favour of dust intrinsic to the NLR clouds with 
varying column density along the
line of sight.
Both electron density and ionisation parameter decrease with radius. 
The decreasing electron density argues against a simple Str\"omgren behaviour
of the NLR gas which has been suggested by the NLR size--luminosity relation
with a slope of 0.33.
The decreasing ionisation parameter in this individual object
does not rule out a common effective ionisation parameter in AGNs as suggested
by  a slope of 0.5 of the NLR size--luminosity relation. 
In the outer part of the NLR, the secondary peak in the electron density and
ionisation--parameter distribution are interpreted as signs of shocks, a
scenario also supported by other observations.

The NLR and stellar velocity fields are similar and indicate that
the NLR gas is distributed in a disk rather than a sphere.
The differences between the two velocity fields can be due to 
several effects such as a rotating gas disk or ring,
gas moving along elliptical streamlines in a barred potential
or outflowing gas from a jet interaction.

The methods presented here were applied to a larger sample of Seyfert
 galaxies, revealing similar results to that of \object{NGC\,1386}. 
A detailed discussion can be
found in \citet{ben05} and will be summarised in an upcoming paper.
\begin{acknowledgements}
N.B. is grateful for financial support by the ``Studienstiftung
des deutschen Volkes''. B.J. acknowledges the support of the Research 
Training Network ``Euro3D--Promoting 3D Spectroscopy in Europe''
(European Commission, Human Potential Network Contract No. 
HPRN--CT--2002--00305) and of the Czech Science Foundation
(grant No. 202/01/D075). M.H. is supported by ``Nordrhein--Westf\"alische
Akademie der Wissenschaften''.
We thank Pierre Ferruit for providing and helping
us with the \texttt{fit/spec} line--fitting tool and Gary Ferland for providing \texttt{CLOUDY}.
Henrique Schmitt was so kind to provide the continuum--subtracted
HST [\ion{O}{iii}] image of \object{NGC\,1386}.
This research has made use of the NASA/IPAC Extragalactic Database (NED), 
operated by the Jet Propulsion Laboratory, Caltech, under contract with the NASA.
\end{acknowledgements}

\appendix
\section{Influence of stellar template}
\label{stellar}
To probe the influence of 
the stellar template used, we
carried out two approaches of (i)
removing the weak emission lines of [\ion{O}{ii}], [\ion{O}{iii}], and
[\ion{N}{ii}] still visible in the template
and (ii) not using any template at all.

If we remove the weak emission lines in the stellar template (method i), the resulting
line ratios are the same within the errors for most spectra, 
i.e. the weak emission is negligible
against the strong NLR emission. The exceptions are the line ratios of
two outer spectra (H, I)
which are shifted towards higher [\ion{N}{ii}]/H$\alpha$ values and
now lie at the edge between \ion{H}{ii} region and LINER region.
This is a possible  explanation for the discrepancy between our results and that of
\citet{wea91} who find LINER--type line ratios in the outer region.
They modified their stellar template derived from the galaxy by removing
the remaining
weak [\ion{N}{ii}] and H$\alpha$ emission and replacing H$\alpha$ by an
artificial absorption line. 
This is a more physical approach than method (i), i.e. 
not only removing the remaining weak emission, but at the same time
deepening the H$\beta$ and H$\alpha$
absorption trough. The true absorption will probably be 
underestimated due to weak
emission in these lines. However, such a correction is difficult to estimate
and the H$\alpha$ absorption used by \citet{wea91} probably still
underestimates the true absorption. 

We decided to not remove the weak emission:
On the one hand, increasing the observed 
[\ion{O}{iii}] and [\ion{N}{ii}] emission by
clipping the weak emission in the template and at the same time
increasing the observed H$\alpha$ and H$\beta$ emission by deepening the
absorption lines in the template will result in approximately the same line
ratios and thus will not have a great influence on the diagnostic diagrams.
On the other hand, the weak emission lines are attributable to ionised regions 
around young stars in
the galaxy lying along our line--of--sight,
and do not originate from
the NLR emission. Thus, they truly contaminate the NLR emission and need
to be subtracted. 

Applying no correction for underlying absorption at all (method ii) yields
extremely large [\ion{O}{iii}]/H$\beta$ values ($\ge$25)
near the nucleus as has already been pointed
out by \citet{wea91}. In the outer parts, the absorption trough
is clearly visible with a small H$\beta$ emission peak in the center. 
In addition, the spectra show galaxy absorption features
which accompany the absorption in H$\beta$. These findings 
show the need for correcting the stellar absorption lines using a suited 
stellar template.

\section{Influence of reddening}
\label{ionired}
The only physical parameter in our study 
which is strongly sensitive to reddening is
the ionisation parameter, derived 
using the line ratio [\ion{O}{ii}]/[\ion{O}{iii}].
We here briefly discuss the influence of the different reddening measures on
the ionisation parameter.

\begin{figure}[h!]
\centering
 \resizebox{\hsize}{!}{\includegraphics[angle=-90]{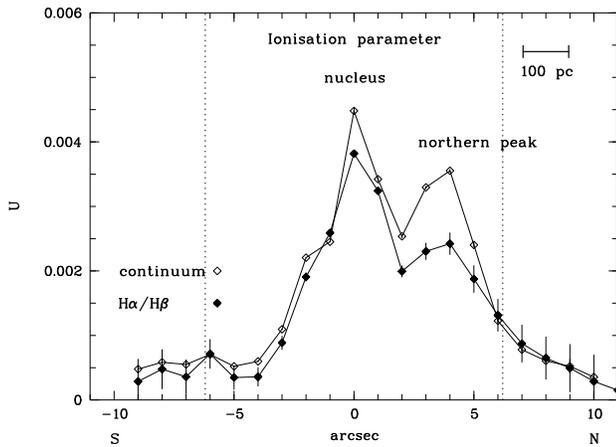}}
\caption{\label{appioni} Ionisation parameter
derived from [\ion{O}{ii}]/[\ion{O}{iii}] ratio
as a function of distance from the nucleus for $n_H$ =
100\,cm$^{-3}$. We compare the ionisation parameter derived 
using the reddening determined from the continuum slope 
for correction (open symbols) with the one from Fig.~\ref{ioni}
using the Balmer decrement to correct for the reddening (filled symbols).
 The edge of the NLR as determined from the
  diagnostic diagrams is indicated by dotted lines. 
}
\end{figure}

Although we corrected for the
reddening within the NLR using the recombination value H$\alpha$/H$\beta$,
there is a clear difference between the reddening
of the continuum slope and that determined from the Balmer decrement:
While the reddening distribution determined by the ratio H$\alpha$/H$\beta$
shows a rather random distribution, the continuum slope reddening
steadily increases towards the nucleus with the exception
of the blue continuum at 3--4\arcsec~to the north of
the nucleus (Fig.~\ref{slopereddening}).
Can these differences explain the observed slope of the ionisation
parameter, i.e. the general decrease with radius and the secondary peak
at that position at which we see a blue continuum?
Is it possible that we do not correct for the 
reddening intrinsic to the NLR gas
by using the reddening value determined by recombination lines but
would the continuum slope be a better tracer for the reddening distribution?
To probe the difference, we used the reddening determined from the continuum
slope to correct the observed
[\ion{O}{ii}]/[\ion{O}{iii}] emission--line ratio.
The resulting ionisation--parameter distribution
changes only slightly showing an even steeper increase towards the center
and a more pronounced secondary peak (Fig.~\ref{appioni}).

Thus, the observed slope in Fig.~\ref{ioni} 
is not an artefact due to a ``wrong'' reddening correction from the
recombination value: Using the reddening determined from the continuum slope, 
the ionisation--parameter distribution
does not flatten but, in contrary, steepen.
The general conclusion of a decreasing ionisation parameter
with radius which shows a secondary peak 3--4\arcsec~north of the nucleus
remains unchanged. However, we consider the recombination value H$\alpha$/H$\beta$ as
a better indicator of the reddening distribution within the NLR
as it uses emission lines originating in the NLR itself. We suggest that the
differences in the reddening measurements originate from 
dust within the NLR with 
varying column density along the
line of sight.
Although in general using the H$\alpha$/H$\beta$ value for reddening
correction bears the problem of the uncertain amount of the underlying stellar
absorption, we believe that our correction using the galaxy itself as stellar
template gives a good approach of the absorption--line free ratio.

\end{document}